\documentclass[journal=nalefd,manuscript=article,layout=twocolumn]{achemso}
\usepackage{graphicx}  
\usepackage{dcolumn}   
\usepackage{bm}        
\usepackage{amssymb}   
\usepackage{amsmath}
\usepackage{color}
\usepackage[colorlinks=true, pdfstartview=FitV, linkcolor=blue, citecolor=blue, urlcolor=blue]{hyperref} 
\usepackage{epstopdf}
\usepackage[free-standing-units=true]{siunitx} 
\usepackage{xpatch}

\usepackage{textcomp}

\DeclareSymbolFont{extraup}{U}{zavm}{m}{n}
\DeclareMathSymbol{\varheart}{\mathalpha}{extraup}{86}

\let\baraccent=\= 

\newcommand{\im}[1]{\text{Im} #1}

\newcommand{\commentOut}[1]{}

\newcommand{\affilIFP}{Laboratory for Solid State Physics, ETH Z\"{u}rich, CH-8093 Z\"urich, Switzerland.}
\newcommand{\affilRow}{Rowland Institute at Harvard, 100 Edwin H. Land Blvd., Cambridge MA 02142, USA.}

\title{Spatially resolved surface dissipation over metal and dielectric substrates}

\author{Martin H\'eritier}\affiliation{\affilIFP}
\author{Raphael Pachlatko}\affiliation{\affilIFP}
\author{Ye Tao}\affiliation{\affilRow}
\author{John M. Abendroth}\affiliation{\affilIFP}
\author{Christian L. Degen}\affiliation{\affilIFP}
\author{Alexander Eichler}\email{eichlera@phys.ethz.ch}\affiliation{\affilIFP}\email{eichlera@phys.ethz.ch}

\date{\today}

\let\oldmaketitle\maketitle
\let\maketitle\relax

\begin{document}
\twocolumn[
\begin{@twocolumnfalse}
\oldmaketitle
\begin{abstract}
We report spatially resolved measurements of static and fluctuating electric fields over conductive (Au) and non-conductive (SiO\textsubscript{2}) surfaces. 
Using an ultrasensitive `nanoladder' cantilever probe to scan over these surfaces at distances of a few tens of nanometers, we record changes in the probe resonance frequency and damping that we associate with static and fluctuating fields, respectively. We find that the two quantities are spatially correlated and of similar magnitude for the two materials. We quantitatively describe the observed effects on the basis of trapped surface charges and dielectric fluctuations in an adsorbate layer. Our results provide direct, spatial evidence for surface dissipation in adsorbates that affects nanomechanical sensors, trapped ions, superconducting resonators, and color centers in diamond.
\end{abstract}
\end{@twocolumnfalse}
]


\paragraph{Introduction --} The last decades have seen rapid progress in the design and operation of devices for quantum applications. Today, we can build highly coherent qubits and resonators in optical, electrical and mechanical media, and interface these platforms with each other to create hybrid systems~\cite{duan_quantum_2010, xiang_hybrid_2013, aspelmeyer_cavity_2014, degen_quantum_2017, bruzewicz_trapped_2019, kjaergaard_superconducting_2020}. Many important advances became possible through a reduction of the critical dimensions to the nanoscale, making the devices more sensitive, faster in their response, and more suitable for dense packaging. However, as dimensions are scaled down, devices become increasingly susceptible to the harmful influence of fluctuating microscopic degrees of freedom. The coherence of trapped ions~\cite{turchette_heating_2000, labaziewicz_temperature_2008, safavi-naini_microscopic_2011, brownnutt_ion-trap_2015, kumph_electric-field_2016} and of superconducting Josephson circuits~\cite{gao_experimental_2008, wang_improving_2009, muller_towards_2019}, for instance, is limited by fluctuating electrical fields. Evidence points to two-level systems in surface oxides and adsorbates as the microscopic origin of these fields~\cite{labaziewicz_temperature_2008, allcock_reduction_2011, hite_100-fold_2012, safavi-naini_influence_2013}. Similar surface effects could also explain the poorly understood non-contact dissipation between closely spaced bodies that presents an obstacle for ultrasensitive scanning force microscopy~\cite{stipe_noncontact_2001, volokitin_noncontact_2003, zurita-sanchez_friction_2004, kuehn_dielectric_2006, volokitin_near-field_2007, yazdanian_dielectric_2008, kisiel_suppression_2011, she_noncontact_2012, den_haan_spin-mediated_2015, de_voogd_dissipation_2017}. Fluctuating electric fields furthermore affect the coherence of Rydberg atoms~\cite{carter_coherent_2013}, color centers in diamond~\cite{kim_decoherence_2015, jamonneau_competition_2016}, and nanomechanical resonators~\cite{tao_permanent_2015, hamoumi_microscopic_2018}.

In spite of the importance of understanding and overcoming these issues, the precise nature of the fluctuating fields is not sufficiently understood and remains the subject of an ongoing debate. This is largely due to experimental challenges. For instance, while atomic force microscopy (AFM) enables direct surface imaging, it usually lacks the sensitivity required to obtain conclusive evidence on non-contact dissipation beyond a few nanometers distance. Trapped ions, which are ideally suited for such investigations due to their high sensitivity, currently do not offer the imaging possibilities and nanoscale resolution of scanning probe methods~\cite{Maiwald_2009}. For this reason, most previous studies relied on power laws of fluctuating forces or fields as a function of temperature, frequency, or distance~\cite{gotsmann_dynamic_2001,stipe_noncontact_2001, kuehn_dielectric_2006, yazdanian_quantifying_2009, saitoh_gigantic_2010, kisiel_suppression_2011, she_noncontact_2012, chiaverini_insensitivity_2014, bruzewicz_measurement_2015, den_haan_spin-mediated_2015, de_voogd_dissipation_2017, sedlacek_evidence_2018, noel_electric-field_2019}. Such studies, however, are difficult to interpret: on the one hand, a single model can produce different power laws at different distances~\cite{brownnutt_ion-trap_2015} or temperatures~\cite{sedlacek_evidence_2018, noel_electric-field_2019}. On the other hand, various microscopic effects can combine to produce complex phenomena that thwart attempts at a simple explanation~\cite{hite_100-fold_2012}.

In this work, we report direct, experimental evidence for the connection between surface dissipation and static variations in the surface potential. To this end, we employ a `nanoladder' scanning force sensor that can detect force noise on the level of $\SI{1}{\atto\newton\per\sqrt{\hertz}}$~\cite{heritier_nanoladder_2018}. This ultrasensitive probe allows us to produce scans of the non-contact dissipation a few tens of \SI{}{\nm} above surfaces. For comparison, we perform measurements over a metal (Au) and a dielectric (SiO\textsubscript{2}). Over both substrates, we find that regions of high and low dissipation correlate with shifts of the mechanical resonance frequency that are attributed to `voltage patches'~\cite{camp_macroscopic_1991, burnham_work-function_1992, rossi_observations_1992, speake_forces_2003, gaillard_method_2006, robertson_kelvin_2006}. We can reproduce our observations using an established model~\cite{yazdanian_dielectric_2008} for electrical field fluctuation in dielectrics together with basic assumptions~\cite{kumph_electric-field_2016}. Our study provides a key to the understanding of surface dissipation effects and a potential route for improving the coherence of many types of quantum devices.

\begin{figure}[h]
\includegraphics[width=\columnwidth]{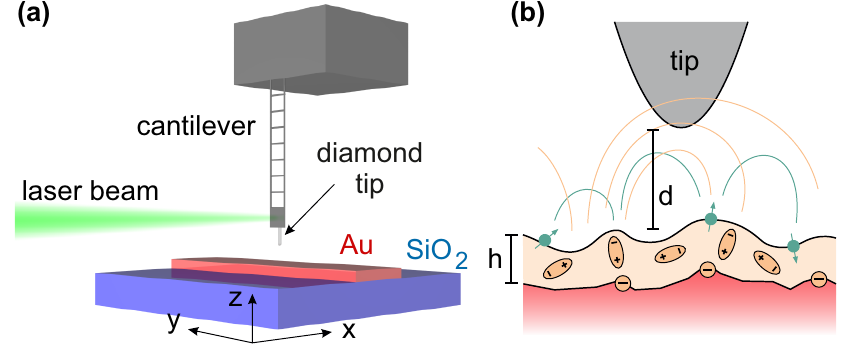}
\caption{\label{fig:fig1} Experimental setup. (a)~A nanoladder cantilever oscillates parallel to the surface in the $x$ direction. It is scanned over a sample surface consisting of SiO\textsubscript{2} and a pattern of Au that is \SI{250}{\nano\meter} thick and electrically grounded. (b)~Schematic representation of the diamond tip interacting with electrical (magnetic) fields generated by charge (spin) defects close to the sample surface. A bright orange area indicates an adsorbant layer.}
\end{figure}

\paragraph{Device and setup.}
Our scanning force probe is a pendulum-style nanoladder cantilever made of single-crystal Si, see Fig.~\ref{fig:fig1}(a)~\cite{heritier_nanoladder_2018}. The cantilever has a bare resonance frequency of $f_\mathrm{0} = \SI{4.858}{\kilo\hertz}$, an effective mass of $m = \SI{2.6}{\pico\gram}$, and a quality factor of $Q_\mathrm{0} = 26100$, corresponding to a spring constant $k_\mathrm{0} = m 4\pi^2f_{\mathrm{0}}^2 = \SI{2.4}{\micro\newton\per\meter}$ and a damping coefficient $\Gamma_\mathrm{0} = m2\pi f_{\mathrm{0}}/Q_\mathrm{0} = \SI{3.1e-15}{\kilo\gram\per\second}$. To obtain a sharp, clean scanning tip, we attach a diamond nanowire~\cite{tao_single_2015, tao_ultrasensitive_2016} to the end of the cantilever with a micro-manipulator. The displacement of the cantilever is detected by fiber-optical interferometry with a $\SI{1550}{\nano\meter}$ laser~\cite{Rugar_1989,heritier_nanoladder_2018}. Measurements are conducted in ultra-high vacuum at a temperature of about $\SI{4}{\kelvin}$. 

\begin{figure}[h!]
  \includegraphics{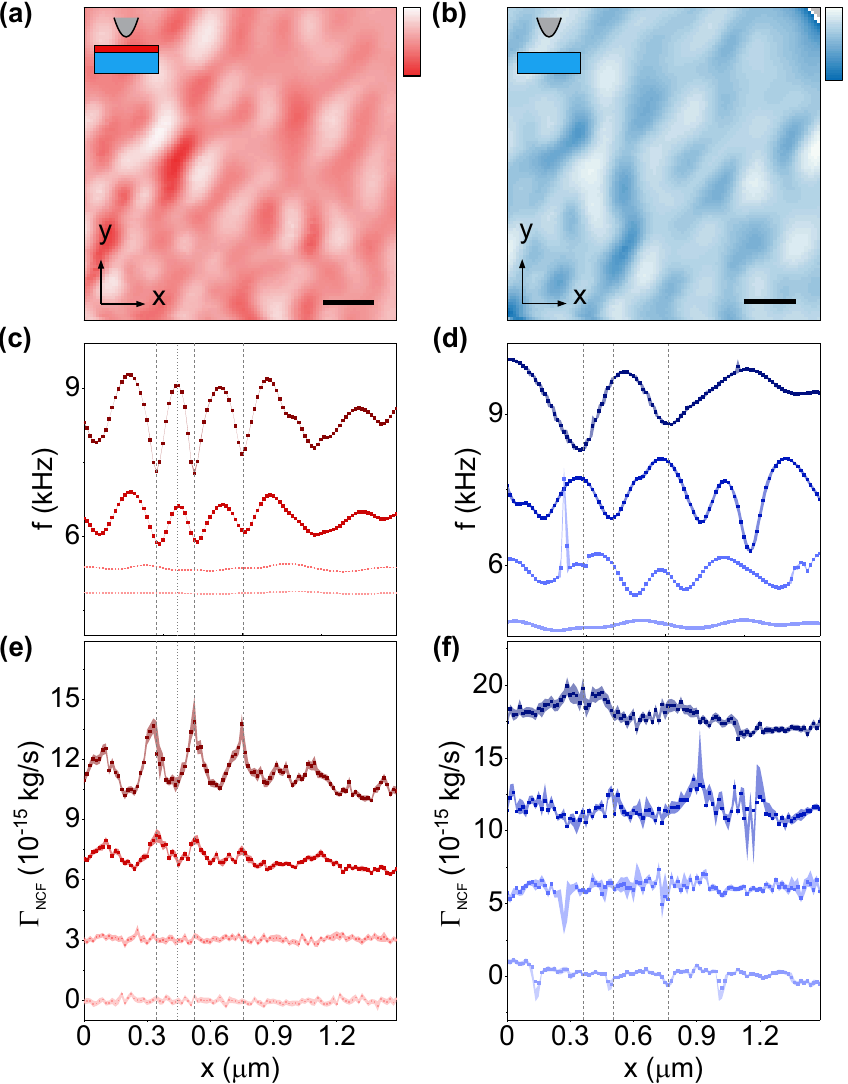}
\caption{\label{fig:fig2} Scanning results. (a)~Maps of the cantilever frequency $f$ recorded at a distance $d=\SI{30}{\nano\meter}$ over Au and (b)~$d=\SI{80}{\nano\meter}$ over SiO\textsubscript{2}. The colors range from red (\SI{4500}{\hertz}) to white (\SI{5000}{\hertz}) in (a) and from blue (\SI{4200}{\hertz}) to white (\SI{5210}{\hertz}) in (b). Both scale bars are \SI{100}{\nano\meter} long. Inserts show the tip over the substrate, SiO\textsubscript{2} (blue) and Au (red). (c)-(f)~Line scans of the resonance frequency $f$ and non-contact friction $\Gamma_{\mathrm{NCF}}$ over Au (c),(e) and SiO\textsubscript{2} (d),(f). The lines corresponds to $d = 150$, $100$, $45$, and $\SI{30}{\nano\meter}$ for Au and $d = 100$, $75$, $60$, $\SI{45}{\nano\meter}$ for SiO\textsubscript{2} (bottom to top). Lines are offset for better visibility by $0$, $0.5$, $1.5$, and $\SI{3.5}{\kilo\hertz}$ in (c), $0$, $1$, $2.5$, and $\SI{4.5}{\kilo\hertz}$ in (d), $\SI{3e-15}{\kilogram\per\second}$ each in (e) and \SI{5e-15}{\kilogram\per\second} each in (f).  Shaded areas denote errors estimated from repeated measurements.}
\end{figure}

\paragraph{Experimental results --}
Our sample surface is a Si substrate with 1500 \SI{}{\nano\meter} of thermally grown SiO\textsubscript{2}. A region of the surface is covered with a \SI{250}{\nano\meter}-thick Au film by e-beam evaporation, see Fig.~\ref{fig:fig1}. We begin our investigations by recording maps of the cantilever frequency $f$ at constant tip height, corresponding to the mean tip-surface distance $d$, see Fig.~\ref{fig:fig2}(a)-(b). The maps reveal distinct, reproducible variations in $f$ on a length scale of ${50}-\SI{150}{\nano\meter}$.
Next, we perform linescans at different values of $d$ and extract $f$ and the quality factor $Q$ from repeated ring-down measurements at every scan position, see Fig.~\ref{fig:fig2}(c)-(d). Data acquisition per point takes $100-\SI{300}{\second}$ for this procedure. From these measurements, we can determine the coefficient of non-contact friction $\Gamma_\mathrm{NCF}$ as
\begin{eqnarray}
    \Gamma_{\mathrm{NCF}} = \Gamma - \Gamma_\mathrm{0} = \frac{2\pi f m}{Q} - \frac{2\pi f_\mathrm{0} m}{Q_\mathrm{0}}\,.
\end{eqnarray}

We observe significant variations of $\Gamma_{\mathrm{NCF}}$ at constant $d$ over both materials, see Fig.~\ref{fig:fig2}(e)-(f). Further, the variations in $\Gamma_{\mathrm{NCF}}$ correlate with those of $f$; maxima of $f$ correspond to minima of $\Gamma_{\mathrm{NCF}}$ and vice versa. The variations smoothen out when increasing $d$. For $d \geq \SI{100}{\nano\meter}$, we retrieve the intrinsic damping of the cantilever, $\Gamma \approx \Gamma_\mathrm{0}$.

\begin{figure}[h!]
\includegraphics[width=\columnwidth]{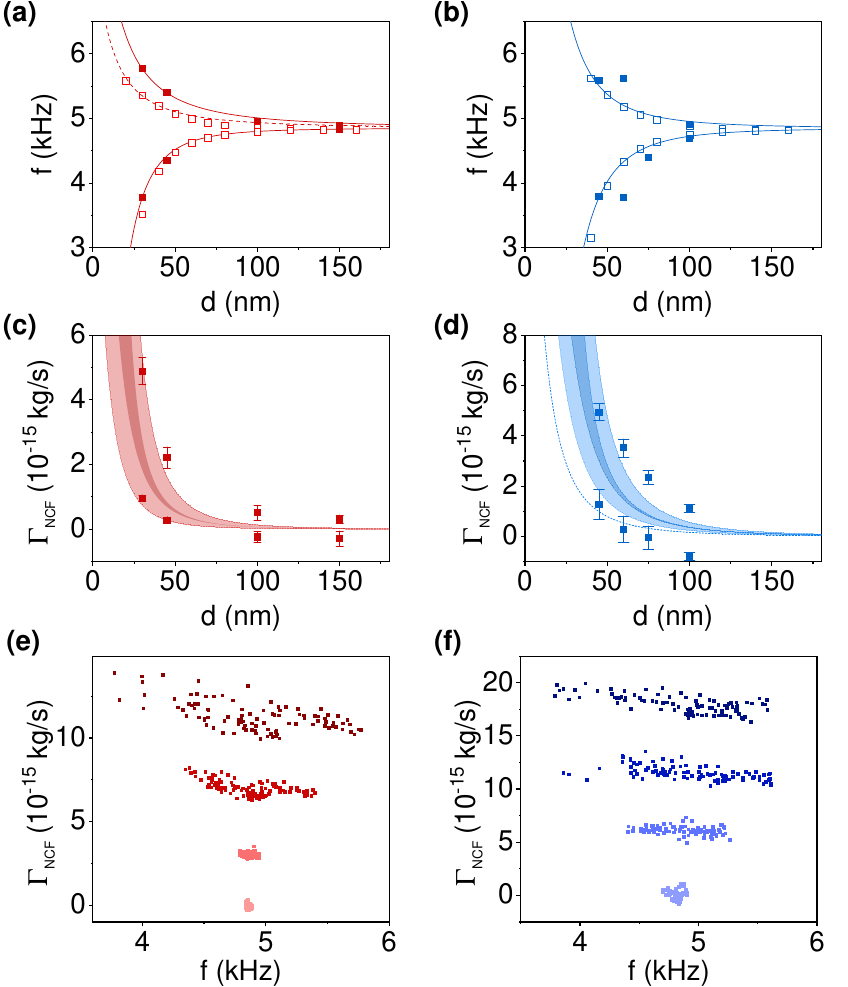}
\caption{\label{fig:fig3} Quantitative data analysis. (a)~Maximum and minimum $f$ as a function of $d$ over Au and (b) over SiO\textsubscript{2}. Filled and open squares correspond to our measurements and to model calculations, respectively. Solid lines are phenomenological fits using $f_0 \pm \beta_\mathrm{\pm} / (d+\Delta)^{\nu_{\pm}}$. For Au, $\beta_- = \SI{3e-23}{\hertz\meter}^{3.5}$ and $\nu_-=3.5$, $\beta_+ = \SI{2.3e-12}{\hertz\square\meter}$ and $\nu_+=2$ ($\beta_+ = \SI{1.25e-12}{\hertz\square\meter}$ for the dashed line). For SiO\textsubscript{2}, $\beta_- = \SI{7.5e-23}{\hertz\meter}^{3.5}$ and $\nu_-=3.5$, $\beta_+ = \SI{1.8e-19}{\hertz\cubic\meter}$ and $\nu_+=3$. See main text and SI for details on the model. (c)~Measured $\Gamma_{\mathrm{NCF}}$ and corresponding model for Au and (d) for SiO\textsubscript{2}. Squares indicate the maxima and minima of a linescan at a distance $d$. The shaded areas corresponds to the model predictions for varying $f$ [cf. solid lines in (a)-(b)] for $h=\SI{1.0}{\nano\meter}$ (dark shade) and for $h$ between $0.4$ and $\SI{2.0}{\nano\meter}$ (bright shade). We use $q_{\mathrm{tip}} = q_e$ for both models, $\epsilon = 2$ and $\tan\theta = 0.01$ for Au, and $\epsilon = 2$ and $\tan\theta = 0.03$ for SiO\textsubscript{2}. The dashed line in (d) is the additive dielectric contribution of the SiO\textsubscript{2} substrate with a thickness of $\SI{1.5}{\micro\meter}$, $\epsilon = 4.44$ and $\tan\theta = 10^{-3}$. (e)~Measured $\Gamma_{\mathrm{NCF}}$ as function of $f$ for different $d$ as in Fig. \ref{fig:fig2} over Au and (f) over SiO\textsubscript{2}. Data sets are offset for better visibility by $\SI{2e-15}{\kilogram\per\second}$ each.}
\end{figure}

Figure~\ref{fig:fig3} summarizes the quantitative analysis of our data. First, we observe a general increase of the variations in both $f$ and $\Gamma_{\mathrm{NCF}}$ with decreasing $d$ over both materials. Second, a strong increase of the mean value of $\Gamma_{\mathrm{NCF}}$ with decreasing $d$ is detected. Third, a correlation between $f$ and $\Gamma_{\mathrm{NCF}}$ is apparent close to the surface.
These features are reproducible over surfaces of Au as well as SiO\textsubscript{2} at different positions over the sample, after thermal cycling to room temperature, and with magnetic fields up to \SI{4}{\tesla} (cf. SI). The results clearly point to non-uniform electric fields generated in a surface layer that must be present without regard of the substrate material underneath. In the following, we discuss concrete models that can explain our measurements.

\paragraph{Model --}
When brought close to a material, the probe tip interacts with electrical or magnetic surface fields, cf. Fig.~\ref{fig:fig1}(b). In general, static fields are expected to modify the cantilever's potential energy $E_{\mathrm{pot}}$, and therefore the spring constant $k = \delta^2 E_{\mathrm{pot}}/\delta x^2$ and the resonance frequency $f = \frac{1}{2\pi}\sqrt{k/m}$~\cite{kozinsky_tuning_2006}. Local variation of the electrical surface potential, dubbed `voltage patches', were previously observed for various materials, and ascribed to trapped charges or work function differences for different crystalline facets~\cite{camp_macroscopic_1991, burnham_work-function_1992, rossi_observations_1992, speake_forces_2003, gaillard_method_2006, robertson_kelvin_2006}. Irrespective of their microscopic origin, we model such voltage patches as isolated charges $q_i$ trapped at the surface. For a point-like tip with a charge $q_{\mathrm{tip}}$, shifts in $f$ can be computed from the added electrostatic potential energy $E_{\mathrm{el}} = \sum_i E_i$, where
\begin{eqnarray}
    E_i = \frac{1}{4\pi \epsilon_0}\frac{q_i q_{\mathrm{tip}}}{r_i}
\end{eqnarray}
is the Coulomb energy of a surface charge $q_i$, with $r_i$ the distance to the cantilever charge $q_{\mathrm{tip}}$. With an electrostatic spring constant $k_{\mathrm{el}} = \delta^2 E_{\mathrm{el}}/\delta x^2$, we obtain
\begin{eqnarray}
    f = \frac{1}{2\pi} \sqrt{\frac{k_\mathrm{0}}{m} + \frac{k_{\mathrm{el}}}{m}} \approx f_0 + \frac{k_\mathrm{el}}{8\pi^2 m f_0}\,,
\end{eqnarray}
where the last term is valid in the limit of $k_\mathrm{el} \ll k_0$.

This simple electrostatic model can reproduce all main features of our frequency scans. The open dots in Fig.~\ref{fig:fig3}(a) show the maximum and minimum frequencies calculated for a square lattice of charges $q_i q_{\mathrm{tip}} = 0.7 q_e^2$, where $q_e = \SI{1.6e-19}{\coulomb}$ is the elementary charge and $0.7$ is an arbitrary scaling factor chosen for best agreement with the experimental data. The model uses a site separation of \SI{150}{\nano\meter} and an offset of $\Delta = \SI{20}{\nano\meter}$ between the tip apex and the position of the effective charge to emulate the shape of the line scans in Fig.~\ref{fig:fig2}(c). The offset roughly corresponds to the expected tip apex radius, cf. SI for details. For SiO\textsubscript{2}, the same model with $q_i q_{\mathrm{tip}} = 1.6 q_e^2$ yields best agreement with the experiment, see Fig.~\ref{fig:fig3}(b). The measured and simulated $f_\mathrm{min,max}$ can be described by phenomenological power laws as described in the figure caption. The model is slightly asymmetric with respect to $f_0$ owing to the difference between the condition for negative $f$ shifts (directly over a charge) and positive shifts (far from charges). This difference can be observed in the experiment as well, for instance for $d=\SI{30}{\nano\meter}$ in Fig.~\ref{fig:fig2}(c). The asymmetry appears to be weaker in the experimental data than in the model, which may be due to a spatial spread of the effective charges in the tip.



While static field gradients give rise to frequency shifts, dissipation is identified as a signature of fluctuating fields. Previous studies have addressed the role of fluctuating electrical or magnetic defects at surfaces for dissipation, both in the contexts of scanning force microscopy~\cite{volokitin_noncontact_2003, zurita-sanchez_friction_2004, kuehn_dielectric_2006, volokitin_near-field_2007, yazdanian_dielectric_2008, she_noncontact_2012, den_haan_spin-mediated_2015, de_voogd_dissipation_2017} and, with a very similar framework, for trapped ions~\cite{turchette_heating_2000, labaziewicz_temperature_2008, safavi-naini_microscopic_2011, brownnutt_ion-trap_2015, kumph_electric-field_2016}. These results established that electrical fluctuations intrinsic to the substrate, such as thermally excited currents or tip-induced mirror charges in a conductor, produce negligible effects under most circumstances~\cite{stipe_noncontact_2001, volokitin_noncontact_2003,zurita-sanchez_friction_2004,kumph_electric-field_2016}. An alternative source of electrical fluctuations could be attributed to thin layers of adsorbants, such as hydrocarbons, that cover a surface immediately upon exposure to air~\cite{smith_hydrophilic_1980,rossi_observations_1992,kuehn_dielectric_2006,kumph_electric-field_2016}. As our cryogenic system does not permit baking out of the sample chamber, we must assume such adsorbant layers to be present. The most basic model for understanding $\Gamma_{\mathrm{NCF}}$, therefore, is based on thermal dielectric fluctuations in a thin layer covering the sample surfaces. We verified with additional measurements that the dominant contribution to $\Gamma_{\mathrm{NCF}}$ cannot be assigned to fluctuating surface electron spins, cf. SI~\cite{she_noncontact_2012}.

We use the model derived in~\cite{yazdanian_dielectric_2008} to determine the value of $\Gamma_{\mathrm{NCF}}$ expected for a thin dielectric. For the adsorbate layer on Au, we use as typical values a relative permittivity $\epsilon = 2$ and a loss tangent $\tan \theta = 0.01$~\cite{kaye_tables_1995,kumph_electric-field_2016}. Defining the complex permittivity $\epsilon_c$ as
\begin{align}
\epsilon_c = \epsilon (1 + i \tan \theta)\,,
\end{align}
as well as the functions
\begin{align}
\zeta = \frac{\epsilon_c-1}{\epsilon_c+1}
\end{align}
\begin{align}
    J_2 = \int_0^\infty \frac{\left(1-\mathrm{e}^{-4u(h/d)}\right) u^2 \mathrm{e}^{-2u} du}{\left( 1 +\zeta'\mathrm{e}^{-2u(h/d)}\right)^2 + \left( \zeta''\mathrm{e}^{-2u(h/d)}\right)^2}
\end{align}where $'$ and $''$ denote real and imaginary parts, respectively, the dissipation is calculated as~\cite{yazdanian_dielectric_2008}
\begin{equation}\label{eq:NCF}
    \Gamma_{\mathrm{NCF}} = \frac{q_{\mathrm{tip}}^2\zeta''}{8\pi^2\epsilon_0 f d^3}J_2
\end{equation}
with $\epsilon_0$ being the permittivity of free space. Note that Eq.~\eqref{eq:NCF} describes the situation of a dielectric on a metal substrate, but we use it also to approximate the adsorbate layer on SiO\textsubscript{2}. In the SI, we present a comparison to a second model that depicts the situation of two nonconducting layers and leads to very similar results~\cite{Lekkala_2013}.

Neither the effective tip charge $q_{\mathrm{tip}}$ nor its exact position in the diamond lattice is controlled in our experiment. The offset of $\Delta=\SI{20}{\nano\meter}$ between the tip apex and the charge position that we introduced for the electrostatic model is included for the dissipation calculations as well. From a rough experimental calibration of the effective tip charge, we get an upper bound of about $20 q_e$ (cf. SI). However, the relevant number of charges must be significantly lower, because the calibration is sensitive to charges on distances of several \SI{}{\micro\meter}, while our experiments only probe interactions on a scale of $d < \SI{100}{nm}$, cf. Fig.~\ref{fig:fig3}(a)-(d). We obtain best results assuming $q_{\mathrm{tip}} \approx q_e$.

There are several ways how Eq.~\eqref{eq:NCF} can be used to explain the experimentally observed variation in $\Gamma_{\mathrm{NCF}}$. We start by noting that Eq.~\eqref{eq:NCF} has an explicit dependency $\Gamma_{\mathrm{NCF}} \propto f^{-1}$, producing the correct trend seen in Fig.~\ref{fig:fig2}. However, the variation of $\Gamma_{\mathrm{NCF}}$ generated in this way is too small to explain our experimental results, see dark shaded area in Fig.~\ref{fig:fig3}(c)-(d), suggesting that additional effects are taking place in parallel. For instance, it was previously found that the thickness of hydrocarbon layers on Au is typically between $h=\SI{0.4}{\nano\meter}$ (a monolayer) and $\SI{2.0}{\nano\meter}$~\cite{Degen_2009,Loretz_2014,kumph_electric-field_2016}. Inserting such a variation in $h$ into Eq.~\eqref{eq:NCF} yields a surprisingly close agreement with our measurements for Au, see Fig.~\ref{fig:fig3}(c). For SiO\textsubscript{2}, the layer thickness required to reproduce our measurement for the given dielectric parameters is about \SI{8}{\nano\meter}, which appears unrealistic. Instead, we present in Fig.~\ref{fig:fig3}(d) a model calculation with the same thickness variation as in (c), but with $\tan \theta = 0.03$. X-ray photoelectron spectroscopy measurements suggest that the chemical composition and bonding nature of the adventitious carbon on the two surfaces is indeed not identical (cf. SI). Considering the open questions regarding dielectric properties of nanometer-scale surface layers, the agreement that we find for both data sets is encouraging. Finally, we also obtain reasonable results when considering variations in $d$. For a thin dielectric layer ($h \ll d$), Eq.~\eqref{eq:NCF} yields approximately $\Gamma_{\mathrm{NCF}} \propto d^{-4}$, resulting in dissipation variations due to sample topography. The surfaces investigated in this work, Au and SiO\textsubscript{2}, show different surface roughness and grain size in AFM topographic scans (see SI). This differences make an interpretation of the $f$ variations in Fig.~\ref{fig:fig2}(a)-(b) in terms of topographic features improbable. Finally, the expected Ohmic loss for bare Au~\cite{kumph_electric-field_2016} turns out to be about ten orders of magnitude smaller than the measured values of $\Gamma_{\mathrm{NCF}}$, ruling out a contribution due to mirror charges in a conductor. In conclusion, it is likely that dielectric fluctuations in a thin surface layer are the dominant cause of non-contact friction in our system.

\paragraph{Discussion --}
We have set out to investigate the surface dissipation over different materials, selecting Au as a representative metal and SiO\textsubscript{2} as a dielectric. Our measurements provide evidence for a correlation between $f$ and $\Gamma_{\mathrm{NCF}}$ over both substrates despite their different electronic properties. The plots in Figs.~\ref{fig:fig3}(e)-(f) suggest $\Gamma_{\mathrm{NCF}} \propto f^{-1}$, even thought the explicit $f^{-1}$ dependency in Eq.~\eqref{eq:NCF}, in concert with the measured variations in $f$, is not sufficient to explain the experimentally observed variations in $\Gamma_{\mathrm{NCF}}$. We propose a simple model where minima (maxima) of $f$ coincide either with maxima (minima) of $h$ or with minima (maxima) of $d$. The microscopic mechanisms behind such correlations are at present speculative, but appear to originate from the electrostatic voltage patches close to the surface. For instance, surface potential patches exert attractive forces onto molecules with a dipole moment, which can lead to site-selective adsorption~\cite{rossi_observations_1992} and a maximum of $h$ directly over static charges. Alternatively, a correlation between topographic features ($d$) and static charges is to be expected if the voltage patches are generated by differences in the surface work function at crystallographic grain orientations, as proposed in Ref.~\cite{gaillard_method_2006}.

With a tip charge estimated as $q_{\mathrm{tip}} = q_e$, we can quantify the power spectral density of the fluctuating electrical field as
\begin{eqnarray}\label{eq:ion_eq}
    S_\mathrm{E} = 4 k_\mathrm{B}T\Gamma_\mathrm{NCF}/q_e^2\,.
\end{eqnarray}
The values we obtain from Eq.~\eqref{eq:ion_eq} are in the range of $\SI{10}{}-\SI{100}{\volt\squared\per\meter\squared\per\hertz}$, which is $10^{11}-10^{15}$ times larger than what is typically detected with trapped ions~\cite{brownnutt_ion-trap_2015}. This discrepancy is not surprising, because the distance to the surface $d$ in our measurements is about $10^3-10^4$ times smaller than in an ion-trap experiment. With the phenomenological power law $\Gamma_{\mathrm{NCF}}\propto d^{-4}$ that we obtained in the thin-film limit from Eq.~\eqref{eq:NCF}, we should expect a difference by a factor $10^{12}-10^{16}$.


It is worth comparing our work to previous studies of fluctuating electrical fields with nanomechanical sensors in the $d = \SI{10}{}-\SI{100}{\nano\meter}$ range~\cite{stipe_noncontact_2001,gotsmann_dynamic_2001,kuehn_dielectric_2006}. We note that those studies concentrated mostly on the $d^{-n}$ dependence at single points over a sample. The values of $n$ that were found varied strongly, from $1<n<1.5$ in Ref.~\cite{stipe_noncontact_2001} to $n \geq 3$ in Ref.~\cite{gotsmann_dynamic_2001}. Our experimental data are in rough agreement with the exponent $n\approx4$ predicted for thin dielectric layers~\cite{kuehn_dielectric_2006,yazdanian_dielectric_2008}. Furthermore, our study also investigates the spatial pattern of $\Gamma_{\mathrm{NCF}}$. The significant variations we find on a \SI{100}{\nano\meter} scale, over conducting and dielectric surfaces alike, offer a potential explanation for the apparent discrepancies in earlier results. In addition, we clearly demonstrate that $\Gamma_{\mathrm{NCF}}$ is connected to electrostatic interactions.

\paragraph{Conclusion and outlook --}
Based on our experimental data and on the agreement with theory, we identify surface adsorbants as the likely origin of non-contact friction over conducting and insulating materials. This result resolves much of the previous disagreement between experiments and models -- even over superconducting surfaces, such thin dielectric layers are often unavoidable. (A notable exception is the experiment from Kisiel et al.~\cite{kisiel_suppression_2011} that was conducted after baking out the vacuum chamber.) A second finding is that fluctuating fields vary spatially and in concert with static surface potentials. This finding provides an important clue to the formation of adsorbants and, potentially, a strategy to reduce their impact. Such a strategy would benefit many of the most advanced fields in quantum sensing and quantum computation, in particular trapped ions, superconducting qubits and ultrasensitive force probes~\cite{poggio_force_2010, brownnutt_ion-trap_2015, kumph_electric-field_2016, degen_quantum_2017, muller_towards_2019}.

In order to reduce $S_\mathrm{E}$ and $\Gamma_\mathrm{NCF}$, future work should focus on the microscopic connection between adsorbants and surface potentials. Nanomechanical probes can be complemented with nuclear magnetic resonance pulses to elucidate the chemical composition of surface layers~\cite{poggio_force_2010, rose_high_2018, grob_magnetic_2019}. Other scanning tools like nanoscale SQUIDS~\cite{finkler_self_2010, vasyukov_imaging_2018} or diamond probes with optically active nitrogen-vacancy defects~\cite{degen_scanning_2008, maze_nanoscale_2008, balasubramanian_nanoscale_2008} could also play a crucial role in the quest to understand and overcome the influence of surface impurities. Finally, directed chemical functionalization with molecular monolayers that feature tailored electronic properties, i.e., defect-tolerant and long-range dipole alignment~\cite{Thomas_2015}, may be explored to mitigate fluctuating electric fields at the interface.

\paragraph{Acknowledgement} We gratefully acknowledge discussions with Jonathan Home,  Thomas Ihn, Marc-Dominik Krass, Roger Loring, John Marohn, Clemens M\"uller, and Oded Zilberberg, the MRFM team of the Degen group, as well as technical support by U. Grob and P. M\"arki. This work is supported by the Swiss National Science Foundation (SNSF) through the Sinergia Project ZEPTO, Grant No. CRSII5$\_$177198/1, the National Center of Competence in Research in Quantum Science and Technology, an ETH Research Grant ETH-03 16-1, and the FIRST cleanroom facility at ETH. Ye Tao is supported by a Rowland Fellowship.

\bibliography{references}

\providecommand{\latin}[1]{#1}
\makeatletter
\providecommand{\doi}
  {\begingroup\let\do\@makeother\dospecials
  \catcode`\{=1 \catcode`\}=2 \doi@aux}
\providecommand{\doi@aux}[1]{\endgroup\texttt{#1}}
\makeatother
\providecommand*\mcitethebibliography{\thebibliography}
\csname @ifundefined\endcsname{endmcitethebibliography}
  {\let\endmcitethebibliography\endthebibliography}{}
\begin{mcitethebibliography}{66}
\providecommand*\natexlab[1]{#1}
\providecommand*\mciteSetBstSublistMode[1]{}
\providecommand*\mciteSetBstMaxWidthForm[2]{}
\providecommand*\mciteBstWouldAddEndPuncttrue
  {\def\EndOfBibitem{\unskip.}}
\providecommand*\mciteBstWouldAddEndPunctfalse
  {\let\EndOfBibitem\relax}
\providecommand*\mciteSetBstMidEndSepPunct[3]{}
\providecommand*\mciteSetBstSublistLabelBeginEnd[3]{}
\providecommand*\EndOfBibitem{}
\mciteSetBstSublistMode{f}
\mciteSetBstMaxWidthForm{subitem}{(\alph{mcitesubitemcount})}
\mciteSetBstSublistLabelBeginEnd
  {\mcitemaxwidthsubitemform\space}
  {\relax}
  {\relax}

\bibitem[Duan and Monroe(2010)Duan, and Monroe]{duan_quantum_2010}
Duan,~L.-M.; Monroe,~C. Colloquium: Quantum networks with trapped ions.
  \emph{Rev. Mod. Phys.} \textbf{2010}, \emph{82}, 1209--1224\relax
\mciteBstWouldAddEndPuncttrue
\mciteSetBstMidEndSepPunct{\mcitedefaultmidpunct}
{\mcitedefaultendpunct}{\mcitedefaultseppunct}\relax
\EndOfBibitem
\bibitem[Xiang \latin{et~al.}(2013)Xiang, Ashhab, You, and
  Nori]{xiang_hybrid_2013}
Xiang,~Z.-L.; Ashhab,~S.; You,~J.~Q.; Nori,~F. Hybrid quantum circuits:
  Superconducting circuits interacting with other quantum systems. \emph{Rev.
  Mod. Phys.} \textbf{2013}, \emph{85}, 623--653\relax
\mciteBstWouldAddEndPuncttrue
\mciteSetBstMidEndSepPunct{\mcitedefaultmidpunct}
{\mcitedefaultendpunct}{\mcitedefaultseppunct}\relax
\EndOfBibitem
\bibitem[Aspelmeyer \latin{et~al.}(2014)Aspelmeyer, Kippenberg, and
  Marquardt]{aspelmeyer_cavity_2014}
Aspelmeyer,~M.; Kippenberg,~T.~J.; Marquardt,~F. Cavity optomechanics.
  \emph{Rev. Mod. Phys.} \textbf{2014}, \emph{86}, 1391--1452\relax
\mciteBstWouldAddEndPuncttrue
\mciteSetBstMidEndSepPunct{\mcitedefaultmidpunct}
{\mcitedefaultendpunct}{\mcitedefaultseppunct}\relax
\EndOfBibitem
\bibitem[Degen \latin{et~al.}(2017)Degen, Reinhard, and
  Cappellaro]{degen_quantum_2017}
Degen,~C.~L.; Reinhard,~F.; Cappellaro,~P. Quantum sensing. \emph{Rev. Mod.
  Phys.} \textbf{2017}, \emph{89}, 035002\relax
\mciteBstWouldAddEndPuncttrue
\mciteSetBstMidEndSepPunct{\mcitedefaultmidpunct}
{\mcitedefaultendpunct}{\mcitedefaultseppunct}\relax
\EndOfBibitem
\bibitem[Bruzewicz \latin{et~al.}(2019)Bruzewicz, Chiaverini, McConnell, and
  Sage]{bruzewicz_trapped_2019}
Bruzewicz,~C.~D.; Chiaverini,~J.; McConnell,~R.; Sage,~J.~M. Trapped-ion
  quantum computing: Progress and challenges. \emph{Applied Physics Reviews}
  \textbf{2019}, \emph{6}, 021314\relax
\mciteBstWouldAddEndPuncttrue
\mciteSetBstMidEndSepPunct{\mcitedefaultmidpunct}
{\mcitedefaultendpunct}{\mcitedefaultseppunct}\relax
\EndOfBibitem
\bibitem[Kjaergaard \latin{et~al.}(2020)Kjaergaard, Schwartz, Braum{\"u}ller,
  Krantz, Wang, Gustavsson, and Oliver]{kjaergaard_superconducting_2020}
Kjaergaard,~M.; Schwartz,~M.~E.; Braum{\"u}ller,~J.; Krantz,~P.; Wang,~J.
  I.-J.; Gustavsson,~S.; Oliver,~W.~D. Superconducting qubits: Current state of
  play. \emph{Annual Review of Condensed Matter Physics} \textbf{2020},
  \emph{11}, 369--395\relax
\mciteBstWouldAddEndPuncttrue
\mciteSetBstMidEndSepPunct{\mcitedefaultmidpunct}
{\mcitedefaultendpunct}{\mcitedefaultseppunct}\relax
\EndOfBibitem
\bibitem[Turchette \latin{et~al.}(2000)Turchette, {Kielpinski}, King,
  Leibfried, Meekhof, Myatt, Rowe, Sackett, Wood, Itano, Monroe, and
  Wineland]{turchette_heating_2000}
Turchette,~Q.~A.; {Kielpinski},; King,~B.~E.; Leibfried,~D.; Meekhof,~D.~M.;
  Myatt,~C.~J.; Rowe,~M.~A.; Sackett,~C.~A.; Wood,~C.~S.; Itano,~W.~M.;
  Monroe,~C.; Wineland,~D.~J. Heating of trapped ions from the quantum ground
  state. \emph{Physical Review A} \textbf{2000}, \emph{61}, 063418, Publisher:
  American Physical Society\relax
\mciteBstWouldAddEndPuncttrue
\mciteSetBstMidEndSepPunct{\mcitedefaultmidpunct}
{\mcitedefaultendpunct}{\mcitedefaultseppunct}\relax
\EndOfBibitem
\bibitem[Labaziewicz \latin{et~al.}(2008)Labaziewicz, Ge, Leibrandt, Wang,
  Shewmon, and Chuang]{labaziewicz_temperature_2008}
Labaziewicz,~J.; Ge,~Y.; Leibrandt,~D.~R.; Wang,~S.~X.; Shewmon,~R.;
  Chuang,~I.~L. Temperature {Dependence} of {Electric} {Field} {Noise} above
  {Gold} {Surfaces}. \emph{Physical Review Letters} \textbf{2008}, \emph{101},
  180602\relax
\mciteBstWouldAddEndPuncttrue
\mciteSetBstMidEndSepPunct{\mcitedefaultmidpunct}
{\mcitedefaultendpunct}{\mcitedefaultseppunct}\relax
\EndOfBibitem
\bibitem[Safavi-Naini \latin{et~al.}(2011)Safavi-Naini, Rabl, Weck, and
  Sadeghpour]{safavi-naini_microscopic_2011}
Safavi-Naini,~A.; Rabl,~P.; Weck,~P.~F.; Sadeghpour,~H.~R. Microscopic model of
  electric-field-noise heating in ion traps. \emph{Physical Review A}
  \textbf{2011}, \emph{84}, 023412, Publisher: American Physical Society\relax
\mciteBstWouldAddEndPuncttrue
\mciteSetBstMidEndSepPunct{\mcitedefaultmidpunct}
{\mcitedefaultendpunct}{\mcitedefaultseppunct}\relax
\EndOfBibitem
\bibitem[Brownnutt \latin{et~al.}(2015)Brownnutt, Kumph, Rabl, and
  Blatt]{brownnutt_ion-trap_2015}
Brownnutt,~M.; Kumph,~M.; Rabl,~P.; Blatt,~R. Ion-trap measurements of
  electric-field noise near surfaces. \emph{Reviews of Modern Physics}
  \textbf{2015}, \emph{87}, 1419--1482, Publisher: American Physical
  Society\relax
\mciteBstWouldAddEndPuncttrue
\mciteSetBstMidEndSepPunct{\mcitedefaultmidpunct}
{\mcitedefaultendpunct}{\mcitedefaultseppunct}\relax
\EndOfBibitem
\bibitem[Kumph \latin{et~al.}(2016)Kumph, Henkel, Rabl, Brownnutt, and
  Blatt]{kumph_electric-field_2016}
Kumph,~M.; Henkel,~C.; Rabl,~P.; Brownnutt,~M.; Blatt,~R. Electric-field noise
  above a thin dielectric layer on metal electrodes. \emph{New Journal of
  Physics} \textbf{2016}, \emph{18}, 023020, Publisher: IOP Publishing\relax
\mciteBstWouldAddEndPuncttrue
\mciteSetBstMidEndSepPunct{\mcitedefaultmidpunct}
{\mcitedefaultendpunct}{\mcitedefaultseppunct}\relax
\EndOfBibitem
\bibitem[Gao \latin{et~al.}(2008)Gao, Daal, Vayonakis, Kumar, Zmuidzinas,
  Sadoulet, Mazin, Day, and Leduc]{gao_experimental_2008}
Gao,~J.; Daal,~M.; Vayonakis,~A.; Kumar,~S.; Zmuidzinas,~J.; Sadoulet,~B.;
  Mazin,~B.~A.; Day,~P.~K.; Leduc,~H.~G. Experimental evidence for a surface
  distribution of two-level systems in superconducting lithographed microwave
  resonators. \emph{Applied Physics Letters} \textbf{2008}, \emph{92},
  152505\relax
\mciteBstWouldAddEndPuncttrue
\mciteSetBstMidEndSepPunct{\mcitedefaultmidpunct}
{\mcitedefaultendpunct}{\mcitedefaultseppunct}\relax
\EndOfBibitem
\bibitem[Wang \latin{et~al.}(2009)Wang, Hofheinz, Wenner, Ansmann, Bialczak,
  Lenander, Lucero, Neeley, O’Connell, Sank, Weides, Cleland, and
  Martinis]{wang_improving_2009}
Wang,~H.; Hofheinz,~M.; Wenner,~J.; Ansmann,~M.; Bialczak,~R.~C.; Lenander,~M.;
  Lucero,~E.; Neeley,~M.; O’Connell,~A.~D.; Sank,~D.; Weides,~M.;
  Cleland,~A.~N.; Martinis,~J.~M. Improving the coherence time of
  superconducting coplanar resonators. \emph{Applied Physics Letters}
  \textbf{2009}, \emph{95}, 233508\relax
\mciteBstWouldAddEndPuncttrue
\mciteSetBstMidEndSepPunct{\mcitedefaultmidpunct}
{\mcitedefaultendpunct}{\mcitedefaultseppunct}\relax
\EndOfBibitem
\bibitem[Müller \latin{et~al.}(2019)Müller, Cole, and
  Lisenfeld]{muller_towards_2019}
Müller,~C.; Cole,~J.~H.; Lisenfeld,~J. Towards understanding two-level-systems
  in amorphous solids: insights from quantum circuits. \emph{Reports on
  Progress in Physics} \textbf{2019}, \emph{82}, 124501, Publisher: IOP
  Publishing\relax
\mciteBstWouldAddEndPuncttrue
\mciteSetBstMidEndSepPunct{\mcitedefaultmidpunct}
{\mcitedefaultendpunct}{\mcitedefaultseppunct}\relax
\EndOfBibitem
\bibitem[Allcock \latin{et~al.}(2011)Allcock, Guidoni, Harty, Ballance, Blain,
  Steane, and Lucas]{allcock_reduction_2011}
Allcock,~D.; Guidoni,~L.; Harty,~T.; Ballance,~C.; Blain,~M.; Steane,~A.;
  Lucas,~D. Reduction of heating rate in a microfabricated ion trap by
  pulsed-laser cleaning. \emph{New Journal of Physics} \textbf{2011},
  \emph{13}, 123023\relax
\mciteBstWouldAddEndPuncttrue
\mciteSetBstMidEndSepPunct{\mcitedefaultmidpunct}
{\mcitedefaultendpunct}{\mcitedefaultseppunct}\relax
\EndOfBibitem
\bibitem[Hite \latin{et~al.}(2012)Hite, Colombe, Wilson, Brown, Warring,
  Jördens, Jost, McKay, Pappas, Leibfried, and Wineland]{hite_100-fold_2012}
Hite,~D.~A.; Colombe,~Y.; Wilson,~A.~C.; Brown,~K.~R.; Warring,~U.;
  Jördens,~R.; Jost,~J.~D.; McKay,~K.~S.; Pappas,~D.~P.; Leibfried,~D.;
  Wineland,~D.~J. 100-{Fold} {Reduction} of {Electric}-{Field} {Noise} in an
  {Ion} {Trap} {Cleaned} with {In} {Situ} {Argon}-{Ion}-{Beam} {Bombardment}.
  \emph{Physical Review Letters} \textbf{2012}, \emph{109}, 103001, Publisher:
  American Physical Society\relax
\mciteBstWouldAddEndPuncttrue
\mciteSetBstMidEndSepPunct{\mcitedefaultmidpunct}
{\mcitedefaultendpunct}{\mcitedefaultseppunct}\relax
\EndOfBibitem
\bibitem[Safavi-Naini \latin{et~al.}(2013)Safavi-Naini, Kim, Weck, Rabl, and
  Sadeghpour]{safavi-naini_influence_2013}
Safavi-Naini,~A.; Kim,~E.; Weck,~P.~F.; Rabl,~P.; Sadeghpour,~H.~R. Influence
  of monolayer contamination on electric-field-noise heating in ion traps.
  \emph{Physical Review A} \textbf{2013}, \emph{87}, 023421, Publisher:
  American Physical Society\relax
\mciteBstWouldAddEndPuncttrue
\mciteSetBstMidEndSepPunct{\mcitedefaultmidpunct}
{\mcitedefaultendpunct}{\mcitedefaultseppunct}\relax
\EndOfBibitem
\bibitem[Stipe \latin{et~al.}(2001)Stipe, Mamin, Stowe, Kenny, and
  Rugar]{stipe_noncontact_2001}
Stipe,~B.~C.; Mamin,~H.~J.; Stowe,~T.~D.; Kenny,~T.~W.; Rugar,~D. Noncontact
  {Friction} and {Force} {Fluctuations} between {Closely} {Spaced} {Bodies}.
  \emph{Physical Review Letters} \textbf{2001}, \emph{87}, 096801, Publisher:
  American Physical Society\relax
\mciteBstWouldAddEndPuncttrue
\mciteSetBstMidEndSepPunct{\mcitedefaultmidpunct}
{\mcitedefaultendpunct}{\mcitedefaultseppunct}\relax
\EndOfBibitem
\bibitem[Volokitin and Persson(2003)Volokitin, and
  Persson]{volokitin_noncontact_2003}
Volokitin,~A.~I.; Persson,~B. N.~J. Noncontact friction between nanostructures.
  \emph{Physical Review B} \textbf{2003}, \emph{68}, 155420\relax
\mciteBstWouldAddEndPuncttrue
\mciteSetBstMidEndSepPunct{\mcitedefaultmidpunct}
{\mcitedefaultendpunct}{\mcitedefaultseppunct}\relax
\EndOfBibitem
\bibitem[Zurita-Sánchez \latin{et~al.}(2004)Zurita-Sánchez, Greffet, and
  Novotny]{zurita-sanchez_friction_2004}
Zurita-Sánchez,~J.~R.; Greffet,~J.-J.; Novotny,~L. Friction forces arising
  from fluctuating thermal fields. \emph{Physical Review A} \textbf{2004},
  \emph{69}, 022902, Publisher: American Physical Society\relax
\mciteBstWouldAddEndPuncttrue
\mciteSetBstMidEndSepPunct{\mcitedefaultmidpunct}
{\mcitedefaultendpunct}{\mcitedefaultseppunct}\relax
\EndOfBibitem
\bibitem[Kuehn \latin{et~al.}(2006)Kuehn, Loring, and
  Marohn]{kuehn_dielectric_2006}
Kuehn,~S.; Loring,~R.~F.; Marohn,~J.~A. Dielectric {Fluctuations} and the
  {Origins} of {Noncontact} {Friction}. \emph{Physical Review Letters}
  \textbf{2006}, \emph{96}, 156103, Publisher: American Physical Society\relax
\mciteBstWouldAddEndPuncttrue
\mciteSetBstMidEndSepPunct{\mcitedefaultmidpunct}
{\mcitedefaultendpunct}{\mcitedefaultseppunct}\relax
\EndOfBibitem
\bibitem[Volokitin and Persson(2007)Volokitin, and
  Persson]{volokitin_near-field_2007}
Volokitin,~A.~I.; Persson,~B. N.~J. Near-field radiative heat transfer and
  noncontact friction. \emph{Reviews of Modern Physics} \textbf{2007},
  \emph{79}, 1291--1329\relax
\mciteBstWouldAddEndPuncttrue
\mciteSetBstMidEndSepPunct{\mcitedefaultmidpunct}
{\mcitedefaultendpunct}{\mcitedefaultseppunct}\relax
\EndOfBibitem
\bibitem[Yazdanian \latin{et~al.}(2008)Yazdanian, Marohn, and
  Loring]{yazdanian_dielectric_2008}
Yazdanian,~S.~M.; Marohn,~J.~A.; Loring,~R.~F. Dielectric fluctuations in force
  microscopy: {Noncontact} friction and frequency jitter. \emph{The Journal of
  Chemical Physics} \textbf{2008}, \emph{128}\relax
\mciteBstWouldAddEndPuncttrue
\mciteSetBstMidEndSepPunct{\mcitedefaultmidpunct}
{\mcitedefaultendpunct}{\mcitedefaultseppunct}\relax
\EndOfBibitem
\bibitem[Kisiel \latin{et~al.}(2011)Kisiel, Gnecco, Gysin, Marot, Rast, and
  Meyer]{kisiel_suppression_2011}
Kisiel,~M.; Gnecco,~E.; Gysin,~U.; Marot,~L.; Rast,~S.; Meyer,~E. Suppression
  of electronic friction on {Nb} films in the superconducting state.
  \emph{Nature Materials} \textbf{2011}, \emph{10}, 119--122, Number: 2
  Publisher: Nature Publishing Group\relax
\mciteBstWouldAddEndPuncttrue
\mciteSetBstMidEndSepPunct{\mcitedefaultmidpunct}
{\mcitedefaultendpunct}{\mcitedefaultseppunct}\relax
\EndOfBibitem
\bibitem[She and Balatsky(2012)She, and Balatsky]{she_noncontact_2012}
She,~J.-H.; Balatsky,~A.~V. Noncontact {Friction} and {Relaxational} {Dynamics}
  of {Surface} {Defects}. \emph{Physical Review Letters} \textbf{2012},
  \emph{108}, 136101, Publisher: American Physical Society\relax
\mciteBstWouldAddEndPuncttrue
\mciteSetBstMidEndSepPunct{\mcitedefaultmidpunct}
{\mcitedefaultendpunct}{\mcitedefaultseppunct}\relax
\EndOfBibitem
\bibitem[den Haan \latin{et~al.}(2015)den Haan, Wagenaar, de~Voogd, Koning, and
  Oosterkamp]{den_haan_spin-mediated_2015}
den Haan,~A. M.~J.; Wagenaar,~J. J.~T.; de~Voogd,~J.~M.; Koning,~G.;
  Oosterkamp,~T.~H. Spin-mediated dissipation and frequency shifts of a
  cantilever at {milliKelvin} temperatures. \emph{Physical Review B}
  \textbf{2015}, \emph{92}, 235441, Publisher: American Physical Society\relax
\mciteBstWouldAddEndPuncttrue
\mciteSetBstMidEndSepPunct{\mcitedefaultmidpunct}
{\mcitedefaultendpunct}{\mcitedefaultseppunct}\relax
\EndOfBibitem
\bibitem[de~Voogd \latin{et~al.}(2017)de~Voogd, Wagenaar, and
  Oosterkamp]{de_voogd_dissipation_2017}
de~Voogd,~J.~M.; Wagenaar,~J. J.~T.; Oosterkamp,~T.~H. Dissipation and
  resonance frequency shift of a resonator magnetically coupled to a
  semiclassical spin. \emph{Scientific Reports} \textbf{2017}, \emph{7}, 42239,
  Number: 1 Publisher: Nature Publishing Group\relax
\mciteBstWouldAddEndPuncttrue
\mciteSetBstMidEndSepPunct{\mcitedefaultmidpunct}
{\mcitedefaultendpunct}{\mcitedefaultseppunct}\relax
\EndOfBibitem
\bibitem[Carter and Martin(2013)Carter, and Martin]{carter_coherent_2013}
Carter,~J.~D.; Martin,~J. D.~D. Coherent manipulation of cold Rydberg atoms
  near the surface of an atom chip. \emph{Phys. Rev. A} \textbf{2013},
  \emph{88}, 043429\relax
\mciteBstWouldAddEndPuncttrue
\mciteSetBstMidEndSepPunct{\mcitedefaultmidpunct}
{\mcitedefaultendpunct}{\mcitedefaultseppunct}\relax
\EndOfBibitem
\bibitem[Kim \latin{et~al.}(2015)Kim, Mamin, Sherwood, Ohno, Awschalom, and
  Rugar]{kim_decoherence_2015}
Kim,~M.; Mamin,~H.~J.; Sherwood,~M.~H.; Ohno,~K.; Awschalom,~D.~D.; Rugar,~D.
  Decoherence of Near-Surface Nitrogen-Vacancy Centers Due to Electric Field
  Noise. \emph{Phys. Rev. Lett.} \textbf{2015}, \emph{115}, 087602\relax
\mciteBstWouldAddEndPuncttrue
\mciteSetBstMidEndSepPunct{\mcitedefaultmidpunct}
{\mcitedefaultendpunct}{\mcitedefaultseppunct}\relax
\EndOfBibitem
\bibitem[Jamonneau \latin{et~al.}(2016)Jamonneau, Lesik, Tetienne, Alvizu,
  Mayer, Dr\'eau, Kosen, Roch, Pezzagna, Meijer, Teraji, Kubo, Bertet, Maze,
  and Jacques]{jamonneau_competition_2016}
Jamonneau,~P.; Lesik,~M.; Tetienne,~J.~P.; Alvizu,~I.; Mayer,~L.; Dr\'eau,~A.;
  Kosen,~S.; Roch,~J.-F.; Pezzagna,~S.; Meijer,~J.; Teraji,~T.; Kubo,~Y.;
  Bertet,~P.; Maze,~J.~R.; Jacques,~V. Competition between electric field and
  magnetic field noise in the decoherence of a single spin in diamond.
  \emph{Phys. Rev. B} \textbf{2016}, \emph{93}, 024305\relax
\mciteBstWouldAddEndPuncttrue
\mciteSetBstMidEndSepPunct{\mcitedefaultmidpunct}
{\mcitedefaultendpunct}{\mcitedefaultseppunct}\relax
\EndOfBibitem
\bibitem[Tao \latin{et~al.}(2015)Tao, Navaretti, Hauert, Grob, Poggio, and
  Degen]{tao_permanent_2015}
Tao,~Y.; Navaretti,~P.; Hauert,~R.; Grob,~U.; Poggio,~M.; Degen,~C.~L.
  Permanent reduction of dissipation in nanomechanical Si resonators by
  chemical surface protection. \emph{Nanotechnology} \textbf{2015}, \emph{26},
  465501\relax
\mciteBstWouldAddEndPuncttrue
\mciteSetBstMidEndSepPunct{\mcitedefaultmidpunct}
{\mcitedefaultendpunct}{\mcitedefaultseppunct}\relax
\EndOfBibitem
\bibitem[Hamoumi \latin{et~al.}(2018)Hamoumi, Allain, Hease, Gil-Santos,
  Morgenroth, G\'erard, Lema\^{\i}tre, Leo, and
  Favero]{hamoumi_microscopic_2018}
Hamoumi,~M.; Allain,~P.~E.; Hease,~W.; Gil-Santos,~E.; Morgenroth,~L.;
  G\'erard,~B.; Lema\^{\i}tre,~A.; Leo,~G.; Favero,~I. Microscopic
  Nanomechanical Dissipation in Gallium Arsenide Resonators. \emph{Phys. Rev.
  Lett.} \textbf{2018}, \emph{120}, 223601\relax
\mciteBstWouldAddEndPuncttrue
\mciteSetBstMidEndSepPunct{\mcitedefaultmidpunct}
{\mcitedefaultendpunct}{\mcitedefaultseppunct}\relax
\EndOfBibitem
\bibitem[Maiwald \latin{et~al.}(2009)Maiwald, Leibfried, Britton, Bergquist,
  Leuchs, and Wineland]{Maiwald_2009}
Maiwald,~R.; Leibfried,~D.; Britton,~J.; Bergquist,~J.~C.; Leuchs,~G.;
  Wineland,~D.~J. Stylus ion trap for enhanced access and sensing. \emph{Nature
  Physics} \textbf{2009}, \emph{5}, 551--554\relax
\mciteBstWouldAddEndPuncttrue
\mciteSetBstMidEndSepPunct{\mcitedefaultmidpunct}
{\mcitedefaultendpunct}{\mcitedefaultseppunct}\relax
\EndOfBibitem
\bibitem[Gotsmann and Fuchs(2001)Gotsmann, and Fuchs]{gotsmann_dynamic_2001}
Gotsmann,~B.; Fuchs,~H. Dynamic {Force} {Spectroscopy} of {Conservative} and
  {Dissipative} {Forces} in an {Al}-{Au}(111) {Tip}-{Sample} {System}.
  \emph{Physical Review Letters} \textbf{2001}, \emph{86}, 2597--2600,
  Publisher: American Physical Society\relax
\mciteBstWouldAddEndPuncttrue
\mciteSetBstMidEndSepPunct{\mcitedefaultmidpunct}
{\mcitedefaultendpunct}{\mcitedefaultseppunct}\relax
\EndOfBibitem
\bibitem[Yazdanian \latin{et~al.}(2009)Yazdanian, Hoepker, Kuehn, Loring, and
  Marohn]{yazdanian_quantifying_2009}
Yazdanian,~S.~M.; Hoepker,~N.; Kuehn,~S.; Loring,~R.~F.; Marohn,~J.~A.
  Quantifying {Electric} {Field} {Gradient} {Fluctuations} over {Polymers}
  {Using} {Ultrasensitive} {Cantilevers}. \emph{Nano Letters} \textbf{2009},
  \emph{9}, 2273--2279, Publisher: American Chemical Society\relax
\mciteBstWouldAddEndPuncttrue
\mciteSetBstMidEndSepPunct{\mcitedefaultmidpunct}
{\mcitedefaultendpunct}{\mcitedefaultseppunct}\relax
\EndOfBibitem
\bibitem[Saitoh \latin{et~al.}(2010)Saitoh, Hayashi, Shibayama, and
  Shirahama]{saitoh_gigantic_2010}
Saitoh,~K.; Hayashi,~K.; Shibayama,~Y.; Shirahama,~K. Gigantic {Maximum} of
  {Nanoscale} {Noncontact} {Friction}. \emph{Physical Review Letters}
  \textbf{2010}, \emph{105}, 236103, Publisher: American Physical Society\relax
\mciteBstWouldAddEndPuncttrue
\mciteSetBstMidEndSepPunct{\mcitedefaultmidpunct}
{\mcitedefaultendpunct}{\mcitedefaultseppunct}\relax
\EndOfBibitem
\bibitem[Chiaverini and Sage(2014)Chiaverini, and
  Sage]{chiaverini_insensitivity_2014}
Chiaverini,~J.; Sage,~J.~M. Insensitivity of the rate of ion motional heating
  to trap-electrode material over a large temperature range. \emph{Phys. Rev.
  A} \textbf{2014}, \emph{89}, 012318\relax
\mciteBstWouldAddEndPuncttrue
\mciteSetBstMidEndSepPunct{\mcitedefaultmidpunct}
{\mcitedefaultendpunct}{\mcitedefaultseppunct}\relax
\EndOfBibitem
\bibitem[Bruzewicz \latin{et~al.}(2015)Bruzewicz, Sage, and
  Chiaverini]{bruzewicz_measurement_2015}
Bruzewicz,~C.~D.; Sage,~J.~M.; Chiaverini,~J. Measurement of ion motional
  heating rates over a range of trap frequencies and temperatures. \emph{Phys.
  Rev. A} \textbf{2015}, \emph{91}, 041402\relax
\mciteBstWouldAddEndPuncttrue
\mciteSetBstMidEndSepPunct{\mcitedefaultmidpunct}
{\mcitedefaultendpunct}{\mcitedefaultseppunct}\relax
\EndOfBibitem
\bibitem[Sedlacek \latin{et~al.}(2018)Sedlacek, Stuart, Slichter, Bruzewicz,
  McConnell, Sage, and Chiaverini]{sedlacek_evidence_2018}
Sedlacek,~J.~A.; Stuart,~J.; Slichter,~D.~H.; Bruzewicz,~C.~D.; McConnell,~R.;
  Sage,~J.~M.; Chiaverini,~J. Evidence for multiple mechanisms underlying
  surface electric-field noise in ion traps. \emph{Physical Review A}
  \textbf{2018}, \emph{98}, 063430, Publisher: American Physical Society\relax
\mciteBstWouldAddEndPuncttrue
\mciteSetBstMidEndSepPunct{\mcitedefaultmidpunct}
{\mcitedefaultendpunct}{\mcitedefaultseppunct}\relax
\EndOfBibitem
\bibitem[Noel \latin{et~al.}(2019)Noel, Berlin-Udi, Matthiesen, Yu, Zhou,
  Lordi, and Häffner]{noel_electric-field_2019}
Noel,~C.; Berlin-Udi,~M.; Matthiesen,~C.; Yu,~J.; Zhou,~Y.; Lordi,~V.;
  Häffner,~H. Electric-field noise from thermally activated fluctuators in a
  surface ion trap. \emph{Physical Review A} \textbf{2019}, \emph{99}, 063427,
  Publisher: American Physical Society\relax
\mciteBstWouldAddEndPuncttrue
\mciteSetBstMidEndSepPunct{\mcitedefaultmidpunct}
{\mcitedefaultendpunct}{\mcitedefaultseppunct}\relax
\EndOfBibitem
\bibitem[H{\ifmmode\acute{e}\else\'{e}\fi}ritier
  \latin{et~al.}(2018)H{\ifmmode\acute{e}\else\'{e}\fi}ritier, Eichler, Pan,
  Grob, Shorubalko, Krass, Tao, and Degen]{heritier_nanoladder_2018}
H{\ifmmode\acute{e}\else\'{e}\fi}ritier,~M.; Eichler,~A.; Pan,~Y.; Grob,~U.;
  Shorubalko,~I.; Krass,~M.~D.; Tao,~Y.; Degen,~C.~L. {Nanoladder Cantilevers
  Made from Diamond and Silicon}. \emph{Nano Lett.} \textbf{2018}, \emph{18},
  1814--1818\relax
\mciteBstWouldAddEndPuncttrue
\mciteSetBstMidEndSepPunct{\mcitedefaultmidpunct}
{\mcitedefaultendpunct}{\mcitedefaultseppunct}\relax
\EndOfBibitem
\bibitem[Camp \latin{et~al.}(1991)Camp, Darling, and
  Brown]{camp_macroscopic_1991}
Camp,~J.~B.; Darling,~T.~W.; Brown,~R.~E. Macroscopic variations of surface
  potentials of conductors. \emph{Journal of Applied Physics} \textbf{1991},
  \emph{69}, 7126--7129\relax
\mciteBstWouldAddEndPuncttrue
\mciteSetBstMidEndSepPunct{\mcitedefaultmidpunct}
{\mcitedefaultendpunct}{\mcitedefaultseppunct}\relax
\EndOfBibitem
\bibitem[Burnham \latin{et~al.}(1992)Burnham, Colton, and
  Pollock]{burnham_work-function_1992}
Burnham,~N.~A.; Colton,~R.~J.; Pollock,~H.~M. Work-function anisotropies as an
  origin of long-range surface forces. \emph{Physical Review Letters}
  \textbf{1992}, \emph{69}, 144--147, Publisher: American Physical
  Society\relax
\mciteBstWouldAddEndPuncttrue
\mciteSetBstMidEndSepPunct{\mcitedefaultmidpunct}
{\mcitedefaultendpunct}{\mcitedefaultseppunct}\relax
\EndOfBibitem
\bibitem[Rossi and Opat(1992)Rossi, and Opat]{rossi_observations_1992}
Rossi,~F.; Opat,~G.~I. Observations of the effects of adsorbates on patch
  potentials. \emph{Journal of Physics D: Applied Physics} \textbf{1992},
  \emph{25}, 1349--1353, Publisher: IOP Publishing\relax
\mciteBstWouldAddEndPuncttrue
\mciteSetBstMidEndSepPunct{\mcitedefaultmidpunct}
{\mcitedefaultendpunct}{\mcitedefaultseppunct}\relax
\EndOfBibitem
\bibitem[Speake and Trenkel(2003)Speake, and Trenkel]{speake_forces_2003}
Speake,~C.~C.; Trenkel,~C. Forces between {Conducting} {Surfaces} due to
  {Spatial} {Variations} of {Surface} {Potential}. \emph{Physical Review
  Letters} \textbf{2003}, \emph{90}, 160403\relax
\mciteBstWouldAddEndPuncttrue
\mciteSetBstMidEndSepPunct{\mcitedefaultmidpunct}
{\mcitedefaultendpunct}{\mcitedefaultseppunct}\relax
\EndOfBibitem
\bibitem[Gaillard \latin{et~al.}(2006)Gaillard, Gros-Jean, Mariolle, Bertin,
  and Bsiesy]{gaillard_method_2006}
Gaillard,~N.; Gros-Jean,~M.; Mariolle,~D.; Bertin,~F.; Bsiesy,~A. Method to
  assess the grain crystallographic orientation with a submicronic spatial
  resolution using {Kelvin} probe force microscope. \emph{Applied Physics
  Letters} \textbf{2006}, \emph{89}, 154101, Publisher: American Institute of
  Physics\relax
\mciteBstWouldAddEndPuncttrue
\mciteSetBstMidEndSepPunct{\mcitedefaultmidpunct}
{\mcitedefaultendpunct}{\mcitedefaultseppunct}\relax
\EndOfBibitem
\bibitem[Robertson \latin{et~al.}(2006)Robertson, Blackwood, Buchman, Byer,
  Camp, Gill, Hanson, Williams, and Zhou]{robertson_kelvin_2006}
Robertson,~N.~A.; Blackwood,~J.~R.; Buchman,~S.; Byer,~R.~L.; Camp,~J.;
  Gill,~D.; Hanson,~J.; Williams,~S.; Zhou,~P. Kelvin probe measurements:
  investigations of the patch effect with applications to {ST}-7 and {LISA}.
  \emph{Classical and Quantum Gravity} \textbf{2006}, \emph{23},
  2665--2680\relax
\mciteBstWouldAddEndPuncttrue
\mciteSetBstMidEndSepPunct{\mcitedefaultmidpunct}
{\mcitedefaultendpunct}{\mcitedefaultseppunct}\relax
\EndOfBibitem
\bibitem[Tao and Degen(2015)Tao, and Degen]{tao_single_2015}
Tao,~Y.; Degen,~C.~L. {Single-Crystal Diamond Nanowire Tips for Ultrasensitive
  Force Microscopy}. \emph{Nano Letters} \textbf{2015}, \emph{15},
  7893--7897\relax
\mciteBstWouldAddEndPuncttrue
\mciteSetBstMidEndSepPunct{\mcitedefaultmidpunct}
{\mcitedefaultendpunct}{\mcitedefaultseppunct}\relax
\EndOfBibitem
\bibitem[Tao \latin{et~al.}(2016)Tao, Eichler, Holzherr, and
  Degen]{tao_ultrasensitive_2016}
Tao,~Y.; Eichler,~A.; Holzherr,~T.; Degen,~C.~L. Ultrasensitive mechanical
  detection of magnetic moment using a commercial disk drive write head.
  \emph{Nature Communications} \textbf{2016}, \emph{7}, 1--8\relax
\mciteBstWouldAddEndPuncttrue
\mciteSetBstMidEndSepPunct{\mcitedefaultmidpunct}
{\mcitedefaultendpunct}{\mcitedefaultseppunct}\relax
\EndOfBibitem
\bibitem[Rugar \latin{et~al.}(1989)Rugar, Mamin, and Guethner]{Rugar_1989}
Rugar,~D.; Mamin,~H.; Guethner,~P. Improved fiber-optic interferometer for
  atomic force microscopy. \emph{Applied Physics Letters} \textbf{1989},
  \emph{55}, 2588--2590\relax
\mciteBstWouldAddEndPuncttrue
\mciteSetBstMidEndSepPunct{\mcitedefaultmidpunct}
{\mcitedefaultendpunct}{\mcitedefaultseppunct}\relax
\EndOfBibitem
\bibitem[Kozinsky \latin{et~al.}(2006)Kozinsky, Postma, Bargatin, and
  Roukes]{kozinsky_tuning_2006}
Kozinsky,~I.; Postma,~H.~C.; Bargatin,~I.; Roukes,~M. Tuning nonlinearity,
  dynamic range, and frequency of nanomechanical resonators. \emph{Applied
  Physics Letters} \textbf{2006}, \emph{88}, 253101\relax
\mciteBstWouldAddEndPuncttrue
\mciteSetBstMidEndSepPunct{\mcitedefaultmidpunct}
{\mcitedefaultendpunct}{\mcitedefaultseppunct}\relax
\EndOfBibitem
\bibitem[Smith(1980)]{smith_hydrophilic_1980}
Smith,~T. {The hydrophilic nature of a clean gold surface}. \emph{Journal of
  Colloid and Interface Science} \textbf{1980}, \emph{75}, 51--55\relax
\mciteBstWouldAddEndPuncttrue
\mciteSetBstMidEndSepPunct{\mcitedefaultmidpunct}
{\mcitedefaultendpunct}{\mcitedefaultseppunct}\relax
\EndOfBibitem
\bibitem[Kaye and Laby(1995)Kaye, and Laby]{kaye_tables_1995}
Kaye,~G. W.~C.; Laby,~T.~H. \emph{{Tables of Physical and Chemical Constants}};
  Longman: Harlow, England, UK, 1995\relax
\mciteBstWouldAddEndPuncttrue
\mciteSetBstMidEndSepPunct{\mcitedefaultmidpunct}
{\mcitedefaultendpunct}{\mcitedefaultseppunct}\relax
\EndOfBibitem
\bibitem[Lekkala \latin{et~al.}(2013)Lekkala, Marohn, and Loring]{Lekkala_2013}
Lekkala,~S.; Marohn,~J.~A.; Loring,~R.~F. Electric force microscopy of
  semiconductors: Theory of cantilever frequency fluctuations and noncontact
  friction. \emph{The Journal of Chemical Physics} \textbf{2013}, \emph{139},
  184702\relax
\mciteBstWouldAddEndPuncttrue
\mciteSetBstMidEndSepPunct{\mcitedefaultmidpunct}
{\mcitedefaultendpunct}{\mcitedefaultseppunct}\relax
\EndOfBibitem
\bibitem[Degen \latin{et~al.}(2009)Degen, Poggio, Mamin, Rettner, and
  Rugar]{Degen_2009}
Degen,~C.; Poggio,~M.; Mamin,~H.; Rettner,~C.; Rugar,~D. Nanoscale magnetic
  resonance imaging. \emph{Proceedings of the National Academy of Sciences}
  \textbf{2009}, \emph{106}, 1313--1317\relax
\mciteBstWouldAddEndPuncttrue
\mciteSetBstMidEndSepPunct{\mcitedefaultmidpunct}
{\mcitedefaultendpunct}{\mcitedefaultseppunct}\relax
\EndOfBibitem
\bibitem[Loretz \latin{et~al.}(2014)Loretz, Pezzagna, Meijer, and
  Degen]{Loretz_2014}
Loretz,~M.; Pezzagna,~S.; Meijer,~J.; Degen,~C. Nanoscale nuclear magnetic
  resonance with a 1.9-nm-deep nitrogen-vacancy sensor. \emph{Applied Physics
  Letters} \textbf{2014}, \emph{104}, 033102\relax
\mciteBstWouldAddEndPuncttrue
\mciteSetBstMidEndSepPunct{\mcitedefaultmidpunct}
{\mcitedefaultendpunct}{\mcitedefaultseppunct}\relax
\EndOfBibitem
\bibitem[Poggio and Degen(2010)Poggio, and Degen]{poggio_force_2010}
Poggio,~M.; Degen,~C.~L. Force-detected nuclear magnetic resonance: recent
  advances and future challenges. \emph{Nanotechnology} \textbf{2010},
  \emph{21}, 342001\relax
\mciteBstWouldAddEndPuncttrue
\mciteSetBstMidEndSepPunct{\mcitedefaultmidpunct}
{\mcitedefaultendpunct}{\mcitedefaultseppunct}\relax
\EndOfBibitem
\bibitem[Rose \latin{et~al.}(2018)Rose, Haas, Chen, Jeon, Lauhon, Cory, and
  Budakian]{rose_high_2018}
Rose,~W.; Haas,~H.; Chen,~A.~Q.; Jeon,~N.; Lauhon,~L.~J.; Cory,~D.~G.;
  Budakian,~R. High-resolution nanoscale solid-state nuclear magnetic resonance
  spectroscopy. \emph{Physical Review X} \textbf{2018}, \emph{8}, 011030\relax
\mciteBstWouldAddEndPuncttrue
\mciteSetBstMidEndSepPunct{\mcitedefaultmidpunct}
{\mcitedefaultendpunct}{\mcitedefaultseppunct}\relax
\EndOfBibitem
\bibitem[Grob \latin{et~al.}(2019)Grob, Krass,
  H{\ifmmode\acute{e}\else\'{e}\fi}ritier, Pachlatko, Rhensius,
  Ko{\ifmmode\check{s}\else\v{s}\fi}ata, Moores, Takahashi, Eichler, and
  Degen]{grob_magnetic_2019}
Grob,~U.; Krass,~M.~D.; H{\ifmmode\acute{e}\else\'{e}\fi}ritier,~M.;
  Pachlatko,~R.; Rhensius,~J.; Ko{\ifmmode\check{s}\else\v{s}\fi}ata,~J.;
  Moores,~B.~A.; Takahashi,~H.; Eichler,~A.; Degen,~C.~L. Magnetic Resonance
  Force Microscopy with a One-Dimensional Resolution of 0.9 Nanometers.
  \emph{Nano Letters} \textbf{2019}, \emph{19}, 7935--7940\relax
\mciteBstWouldAddEndPuncttrue
\mciteSetBstMidEndSepPunct{\mcitedefaultmidpunct}
{\mcitedefaultendpunct}{\mcitedefaultseppunct}\relax
\EndOfBibitem
\bibitem[Finkler \latin{et~al.}(2010)Finkler, Segev, Myasoedov, Rappaport,
  Ne’eman, Vasyukov, Zeldov, Huber, Martin, and Yacoby]{finkler_self_2010}
Finkler,~A.; Segev,~Y.; Myasoedov,~Y.; Rappaport,~M.~L.; Ne’eman,~L.;
  Vasyukov,~D.; Zeldov,~E.; Huber,~M.~E.; Martin,~J.; Yacoby,~A. Self-aligned
  nanoscale SQUID on a tip. \emph{Nano Letters} \textbf{2010}, \emph{10},
  1046--1049\relax
\mciteBstWouldAddEndPuncttrue
\mciteSetBstMidEndSepPunct{\mcitedefaultmidpunct}
{\mcitedefaultendpunct}{\mcitedefaultseppunct}\relax
\EndOfBibitem
\bibitem[Vasyukov \latin{et~al.}(2018)Vasyukov, Ceccarelli, Wyss, Gross,
  Schwarb, Mehlin, Rossi, T\"{u}t\"{u}ncu\"{o}glu, Heimbach, Zamani, Kov\'acs,
  Fontcuberta~i Morral, Grundler, and Poggio]{vasyukov_imaging_2018}
Vasyukov,~D.; Ceccarelli,~L.; Wyss,~M.; Gross,~B.; Schwarb,~A.; Mehlin,~A.;
  Rossi,~N.; T\"{u}t\"{u}ncu\"{o}glu,~G.; Heimbach,~F.; Zamani,~R.;
  Kov\'acs,~A.; Fontcuberta~i Morral,~A.; Grundler,~D.; Poggio,~M. Imaging
  stray magnetic field of individual ferromagnetic nanotubes. \emph{Nano
  Letters} \textbf{2018}, \emph{18}, 964--970\relax
\mciteBstWouldAddEndPuncttrue
\mciteSetBstMidEndSepPunct{\mcitedefaultmidpunct}
{\mcitedefaultendpunct}{\mcitedefaultseppunct}\relax
\EndOfBibitem
\bibitem[Degen(2008)]{degen_scanning_2008}
Degen,~C. Scanning magnetic field microscope with a diamond single-spin sensor.
  \emph{Applied Physics Letters} \textbf{2008}, \emph{92}, 243111\relax
\mciteBstWouldAddEndPuncttrue
\mciteSetBstMidEndSepPunct{\mcitedefaultmidpunct}
{\mcitedefaultendpunct}{\mcitedefaultseppunct}\relax
\EndOfBibitem
\bibitem[Maze \latin{et~al.}(2008)Maze, Stanwix, Hodges, Hong, Taylor,
  Cappellaro, Jiang, Dutt, Togan, Zibrov, Yacoby, Walsworth, and
  Lukin]{maze_nanoscale_2008}
Maze,~J.; Stanwix,~P.; Hodges,~J.; Hong,~S.; Taylor,~J.; Cappellaro,~P.;
  Jiang,~L.; Dutt,~M.~G.; Togan,~E.; Zibrov,~A.; Yacoby,~A.; Walsworth,~R.~L.;
  Lukin,~M.~D. Nanoscale magnetic sensing with an individual electronic spin in
  diamond. \emph{Nature} \textbf{2008}, \emph{455}, 644\relax
\mciteBstWouldAddEndPuncttrue
\mciteSetBstMidEndSepPunct{\mcitedefaultmidpunct}
{\mcitedefaultendpunct}{\mcitedefaultseppunct}\relax
\EndOfBibitem
\bibitem[Balasubramanian \latin{et~al.}(2008)Balasubramanian, Chan, Kolesov,
  Al-Hmoud, Tisler, Shin, Kim, Wojcik, Hemmer, Krueger, Hanke, Leitenstorfer,
  Bratschitsch, Jelezko, and Wrachtrup]{balasubramanian_nanoscale_2008}
Balasubramanian,~G.; Chan,~I.; Kolesov,~R.; Al-Hmoud,~M.; Tisler,~J.; Shin,~C.;
  Kim,~C.; Wojcik,~A.; Hemmer,~P.~R.; Krueger,~A.; Hanke,~T.;
  Leitenstorfer,~A.; Bratschitsch,~R.; Jelezko,~F.; Wrachtrup,~J. Nanoscale
  imaging magnetometry with diamond spins under ambient conditions.
  \emph{Nature} \textbf{2008}, \emph{455}, 648\relax
\mciteBstWouldAddEndPuncttrue
\mciteSetBstMidEndSepPunct{\mcitedefaultmidpunct}
{\mcitedefaultendpunct}{\mcitedefaultseppunct}\relax
\EndOfBibitem
\bibitem[Thomas \latin{et~al.}(2015)Thomas, Schwartz, Hohman, Claridge, Auluck,
  Serino, Spokoyny, Tran, Kelly, Mirkin, Gilles, Osher, and Weiss]{Thomas_2015}
Thomas,~J.; Schwartz,~J.; Hohman,~J.; Claridge,~S.; Auluck,~H.; Serino,~A.;
  Spokoyny,~A.; Tran,~G.; Kelly,~K.; Mirkin,~C.; Gilles,~J.; Osher,~S.;
  Weiss,~P. Defect-tolerant aligned dipoles within two-dimensional plastic
  lattices. \emph{ACS Nano} \textbf{2015}, \emph{9}, 4734--4742\relax
\mciteBstWouldAddEndPuncttrue
\mciteSetBstMidEndSepPunct{\mcitedefaultmidpunct}
{\mcitedefaultendpunct}{\mcitedefaultseppunct}\relax
\EndOfBibitem
\end{mcitethebibliography}


\begin{thebibliography}{111}
\raggedright
\bibitem{heritier_nanoladder_2018_s}
M.~H\'{e}ritier, A.~Eichler, Y.~Pan, U.~Grob, I.~Shorubalko, M. D.~Krass, Y.~Tao, and C.~Degen, {\em Nano Letters}, 18, 1814 (2018), ISSN 1530-6984

\bibitem{yazdanian_dielectric_2008_s}
S. M.~Yazdanian,  J. A.~Marohn,  and R. F.~Loring, {\em The Journal of Chemical Physics}, 128 (2008),  ISSN 0021-9606

\bibitem{Lekkala_2013_s}
S.~Lekkala, J. A.~Marohn, and R. F.~Loring, {\em The Journal of chemical physics}, 139, 184702 (2013).

\bibitem{grob_magnetic_2019_s}
U.~Grob, M. D.~Krass, M.~H\'{e}ritier, R.~Pachlatko, J.~Rhensius, J.~Ko\v{o}ata, B. A.~Moores, H.~Takahashi, A.~Eichler, and C. L.~Degen, {\em Nano Letters} 19, 7935 (2019), ISSN 1530-6984

\bibitem{tao_single_2015_s}
Y.~Tao and C. L.~Degen, {\em Nano Letters} 15, 7893 (2015), ISSN 1530-6984

\bibitem{Sen_1992}
A. D.~Sen, V. G.~Anicich, and T.~Arakelian, {\em Journal of Physics D: Applied Physics} 25, 516 (1992)

\bibitem{Sangtawesin2019}
S.~Sangtawesin, B. L.~Dwyer, S.~Srinivasan, J. J.~Allred, L. V. H.~Rodgers, K.~De Greve, A.~Stacey, N.~Dontschuk, K. M.~O’Donnell, D.~Hu, et al., {\em Phys. Rev. X} 9, 031052 (2019)

\end{thebibliography}


\clearpage
\onecolumn

\setlength{\parindent}{1em}
\setlength{\parskip}{.3em}

\large
\begin{center}
\textbf{\Large \titlefont{Supplementary Material: Spatially resolved surface dissipation over metal and dielectric substrates}}

\vspace{5 mm}

{\large Martin H\'{e}ritier$^{\dagger}$, Raphael Pachlatko$^{\dagger}$, Ye Tao$^{\ddagger}$, John M. Abendroth$^{\dagger}$, Christian L. Degen$^{\dagger}$, and Alexander Eichler$^{\dagger}$}\\
\vspace{5 mm}
\normalsize

\textit{$^\dagger$Laboratory for Solid State Physics, ETH Z\"{u}rich, CH-8093 Z\"urich, Switzerland.}\\
 \textit{$^\ddagger$Rowland Institute at Harvard, 100 Edwin H. Land Blvd., Cambridge MA 02142, USA}\\

\end{center}

\makeatletter
\renewcommand{\theequation}{S\arabic{equation}}
\renewcommand{\thefigure}{S\arabic{figure}}
\renewcommand{\thetable}{S\arabic{table}}
\renewcommand{\bibnumfmt}[1]{[S#1]}
\renewcommand{\citenumfont}[1]{S#1}
\setcounter{equation}{0}
\setcounter{figure}{0}
\setcounter{table}{0}
\renewcommand{\figurename}{\textbf{Supplementary Figure}}
\renewcommand{\tablename}{\textbf{Supplementary Table}}

\normalsize
\section{Cantilever characterization}
\begin{figure}[h!]
\centering
\includegraphics{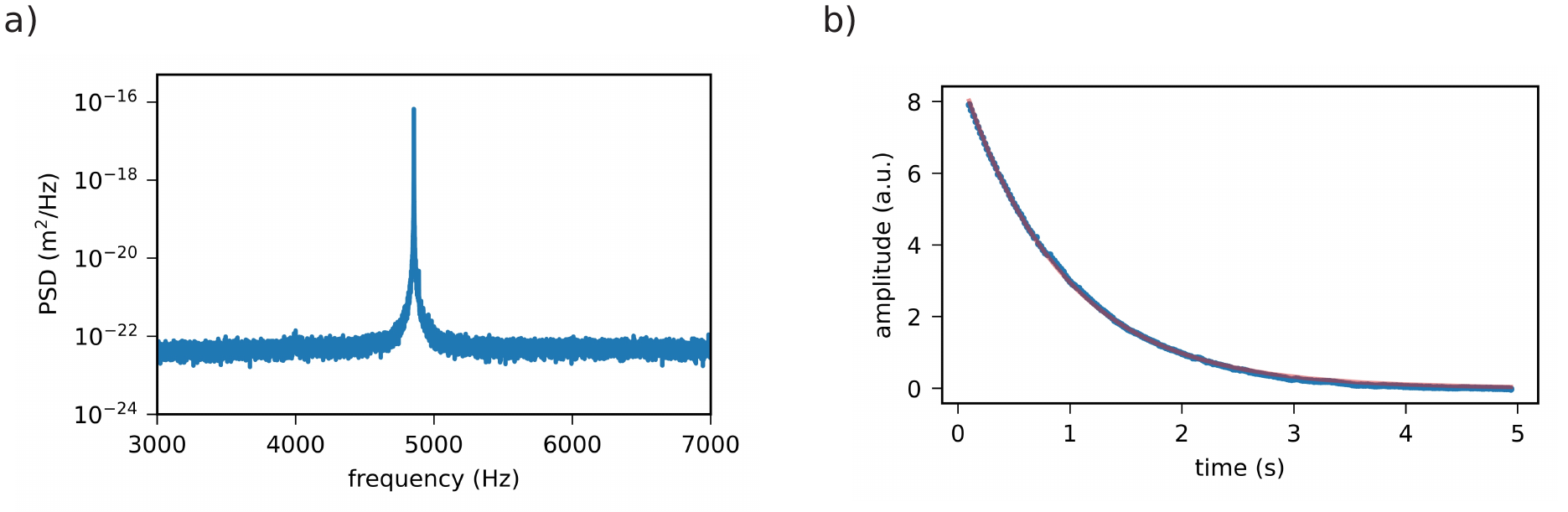}
\caption{(a)~Cantilever displacement power spectral density around its fundamental frequency of $\SI{4.858}{\kilo\hertz}$ at $\SI{4}{\kelvin}$. (b)~Typical cantilever ringdown measurement. The decay of the amplitude is averaged over $10-20$ runs and fitted by an exponential decay (red line).}
\label{fig:cantilevercharacterisation}
\end{figure}

Details about the silicon nanoladder cantilever design were reported in a previous publication~\cite{heritier_nanoladder_2018_s}. In Fig.~\ref{fig:cantilevercharacterisation}(a), we show the displacement power spectral density (PSD), measured at $\SI{4}{\kelvin}$, of the cantilever used in this study. We use such measurements at room temperature to calibrate the effective resonator mass $m$, and repeat them at cryogenic temperatures to verify the effective temperature $T$ of the sensor. A typical ringdown experiment used to determine the $Q$-factor is displayed in Fig.~\ref{fig:cantilevercharacterisation}(b).

\section{Model of two dielectric layers}

In the main text, we use the model by Yazdanian et al. to calculate the non-contact friction due to dielectric fluctuations in thin adsorbant layers on Au and SiO\textsubscript{2}~\cite{yazdanian_dielectric_2008_s}. The model assumes that the dielectric layer is directly supported by a metal, as is the case for our Au sample. For SiO\textsubscript{2}, the situation is slightly different, as the adsorbant layer is supported by a second dielectric (SiO\textsubscript{2}) whose influence we account for by an additive term in $\Gamma_\mathrm{NCF}$ that we calculate with the same equation, see dashed line in Fig.~3(c). We neglect the influence of the Si substrate that is more than \SI{1}{\micro\meter} away from the tip.

A more rigorous way to calculate $\Gamma_\mathrm{NCF}$ for our SiO\textsubscript{2} sample is provided by the model by Lekkala et al.~\cite{Lekkala_2013_s}. In this model, a dielectric and a semiconducting layer are considered. By setting the density of free charges in the semiconductor to be zero, we arrive at a model for two dielectric layers.

\begin{figure}[h!]
	\includegraphics{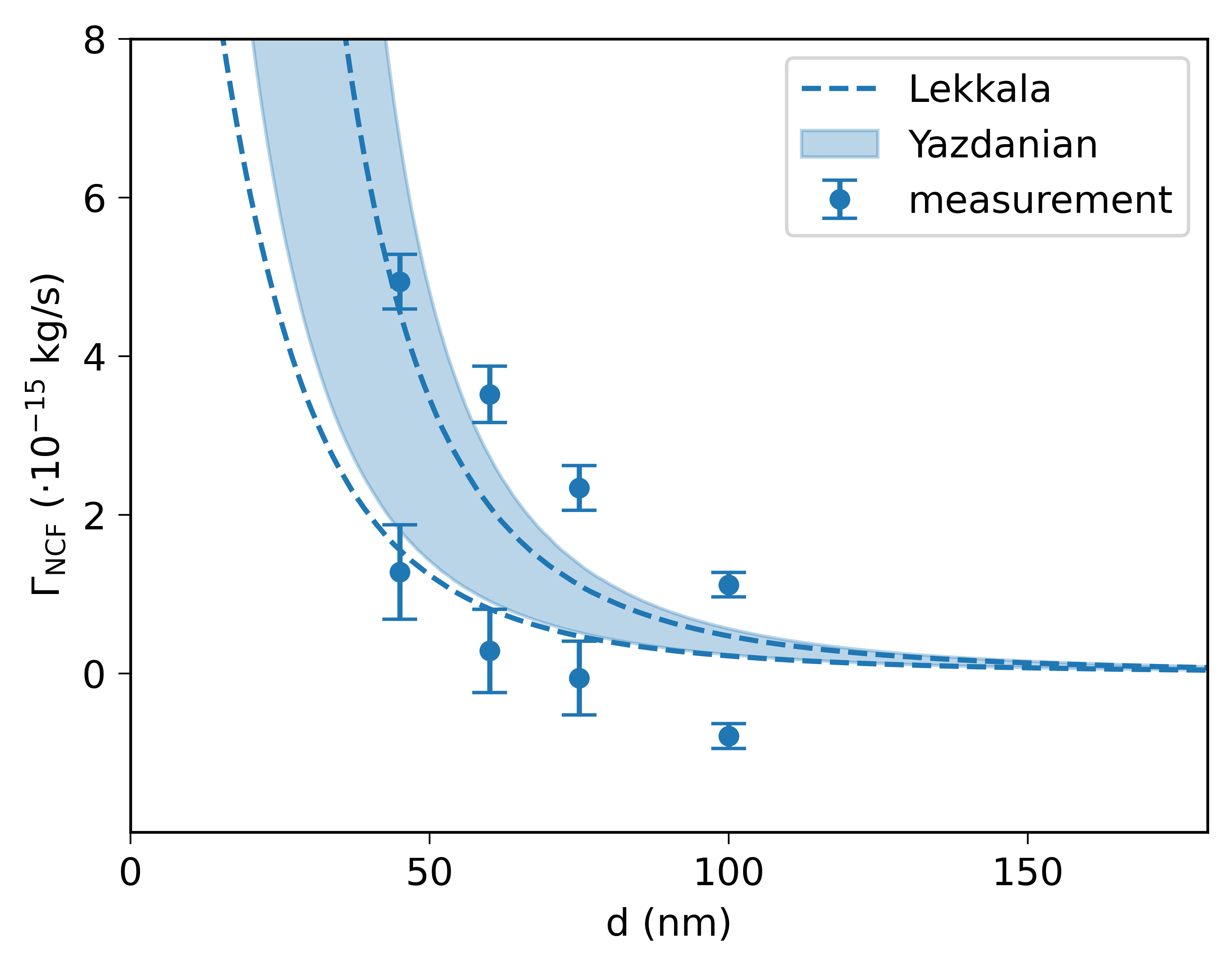}
	\caption{Comparison of the results for $\Gamma_\mathrm{NCF}$ of a dielectric layer with $\epsilon = 2$, $\tan\theta = 0.03$ and $0.4<h<\SI{2}{\nano\meter}$, over a \SI{1.5}{\micro\meter}-thick layer of SiO\textsubscript{2}, calculated with the models by Yazdanian et al.~\cite{yazdanian_dielectric_2008_s} and Lekkala et al.~\cite{Lekkala_2013_s}.}
	\label{fig:lekkala}
\end{figure}

The equation to calculate $\Gamma_\mathrm{NCF}$ with the Lekkala model is~\cite{Lekkala_2013_s}
\begin{align}
   \Gamma_\mathrm{NCF} = -\frac{q_\mathrm{tip}^2}{16\pi^2 f\epsilon_0} \int_0^\infty du u^2 e^{-2ud} \im\left( \frac{\epsilon_c - \xi(u)}{\epsilon_c + \xi(u)} \right)\,
\end{align}
Here, we use the definitions
\begin{align}
   \xi(u) &= \frac{\sinh(uh)^2 + \alpha \cosh(uh) \sinh(uh)}{\alpha \sinh(uh)^2 + \cosh(uh) \sinh(uh)}\,, \\
   \alpha &= \frac{\epsilon_c}{\epsilon_\mathrm{SiO\textsubscript{2}}}\,,
\end{align}
where $\epsilon_\mathrm{SiO\textsubscript{2}}$ is the relative dielectric permittivity of the SiO\textsubscript{2} substrate and all other parameters are defined as in the main text. The results we obtain with this model, using identical values as in the main text, are shown in Fig.~\ref{fig:lekkala}. We find a small quantitative difference compared to the Yazdanian model.

\section{Results with magnetic field}
In this section, we present the data measured under an external magnetic field of $B = \SI{4}{\tesla}$ applied in the z-direction. At this field, the Boltzmann energy $k_{\mathrm{B}} T$ and the magnetic potential energy $\mu_{\mathrm{B}} H$ of electron spins are approximately equal, where $k_{\mathrm{B}} = \SI{1.38e-23}{\joule\per\kelvin}$ is the Boltzmann constant, $T = \SI{4}{\kelvin}$ is the temperature, and $\mu_{\mathrm{B}} = \SI{9.3e-24}{\joule\per\tesla}$ is the Bohr magneton. We would therefore expect a noticeable change in $\Gamma_{\mathrm{NCF}}$ if fluctuating electronic spins are responsible for it.

\begin{figure}[h]
	\centering
	\includegraphics{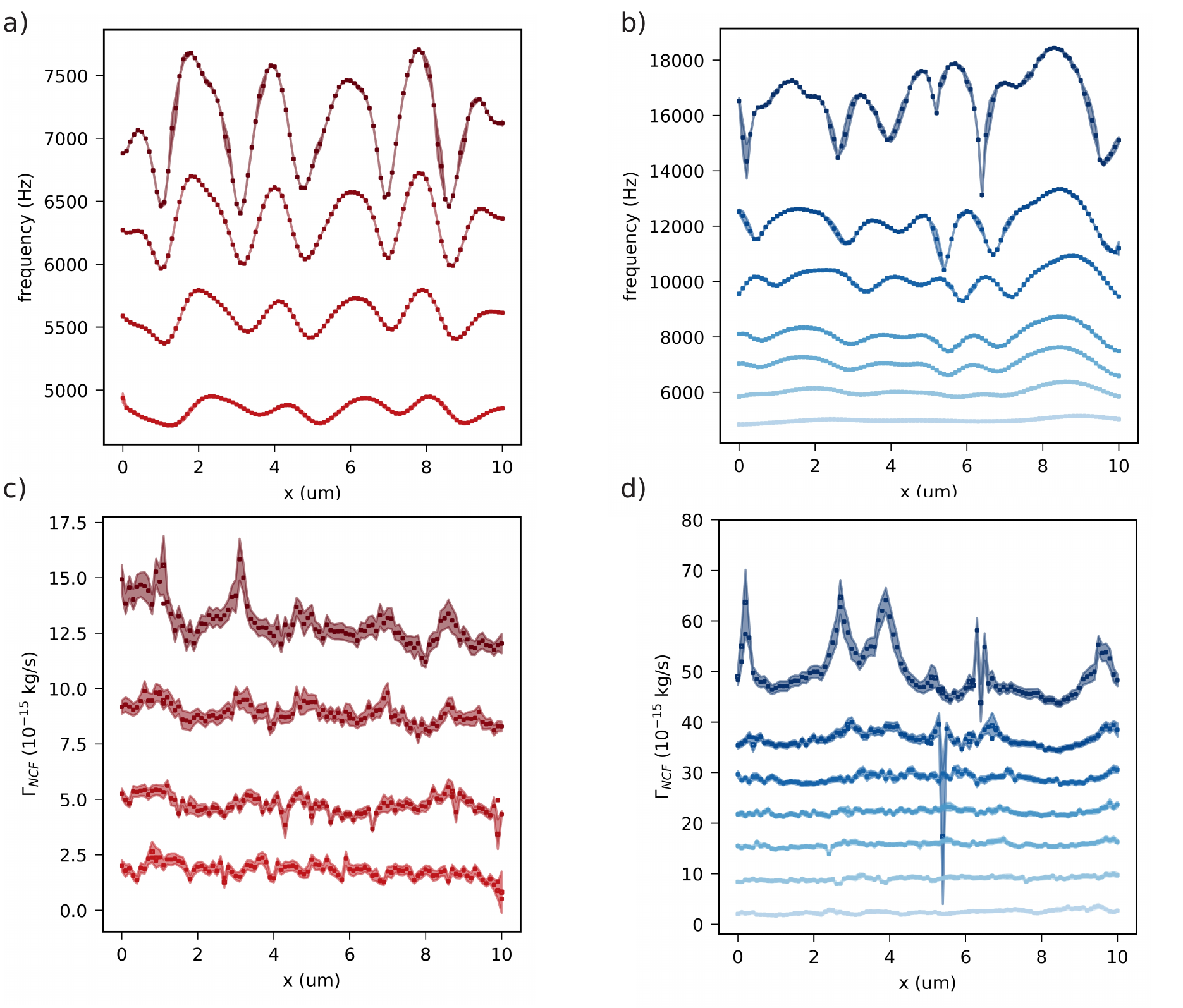}
	\caption{Line scans of the resonance frequency $f$ and non-contact friction $\Gamma_\mathrm{NCF}$ over Au (a)-(c) and SiO\textsubscript{2} (b)-(d) under an external magnetic field of $\SI{4}{\tesla}$ for $d = 20$, $30$, $45$, $\SI{60}{\nano\meter}$ over Au and $d = 30$, $45$, $50$, $70$, $80$, $100$, $\SI{150}{\nano\meter}$ over SiO\textsubscript{2} (top to bottom). Lines are offset for better visibility by $\SI{.075}{\kilo\hertz}$ each in (a), $11$, $7$, $5$, $3$, $2$, $\SI{1}{\kilo\hertz}$ in (b), $\SI{3e-15}{\kilogram\per\second}$ each in (c), and $\SI{6e-15}{\kilogram\per\second}$ each in (d).}
	\label{fig:linescanmagnet}
\end{figure}

\begin{figure}[h]
	\centering
		\includegraphics{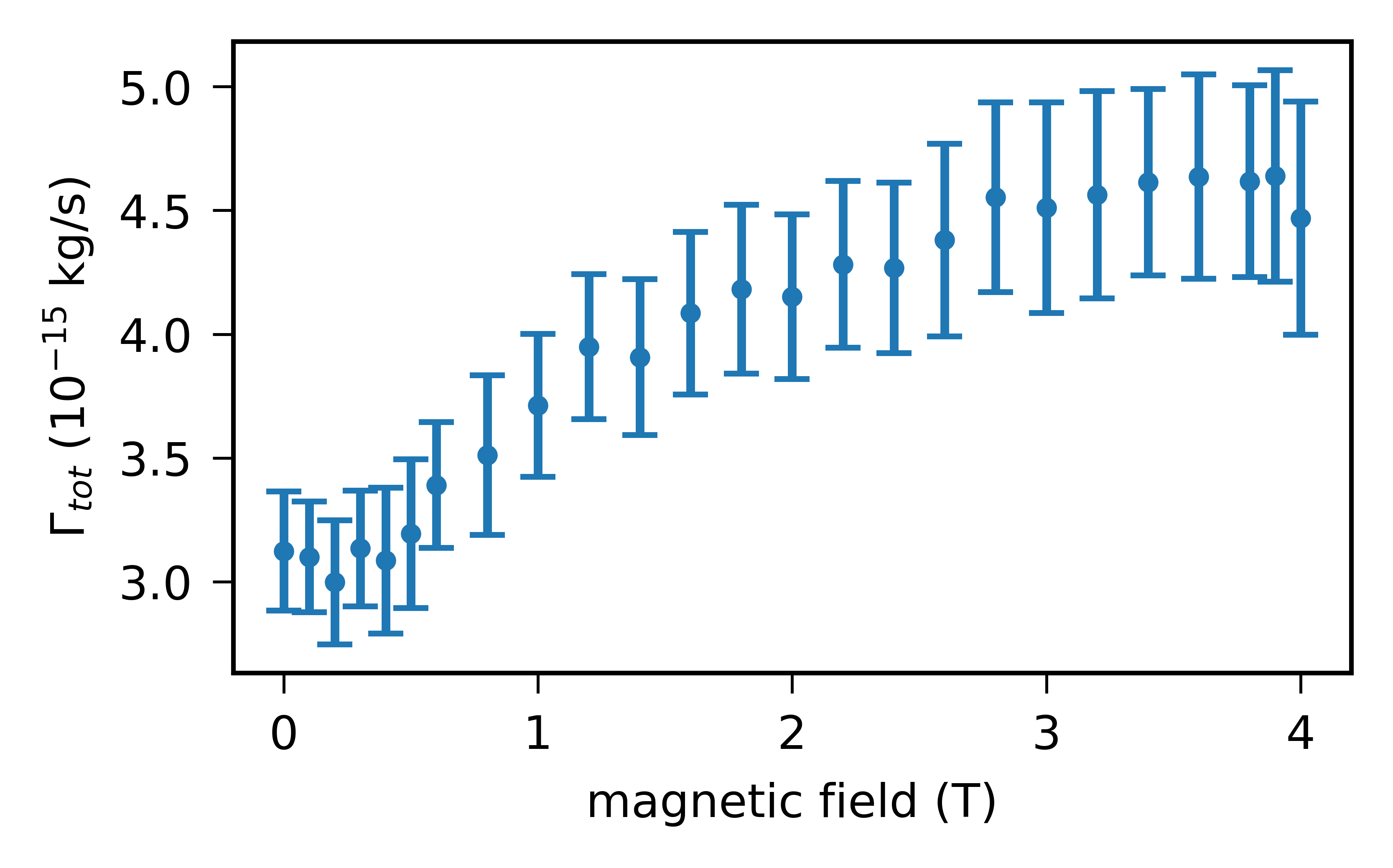}\\[3pt]
	\caption{Dissipation $\Gamma_\mathrm{tot}$ as function of magnetic field far from the surface.}
	\label{fig:magneticfieldapproach}
\end{figure}

In Fig.~\ref{fig:linescanmagnet}, we show the line scans of frequency $f$ and non-contact friction $\Gamma_\mathrm{NCF}$ as in Fig.~2 of the main paper. Figure~\ref{fig:magneticfieldapproach}(a) displays the dependency of the total cantilever dissipation $\Gamma_\mathrm{tot}$ on the external magnetic field far from the surface. We see that the cantilever damping increases with $B$, potentially due to magnetic impurities on the cantilever surface.

\begin{figure}[h]
	\centering
\includegraphics{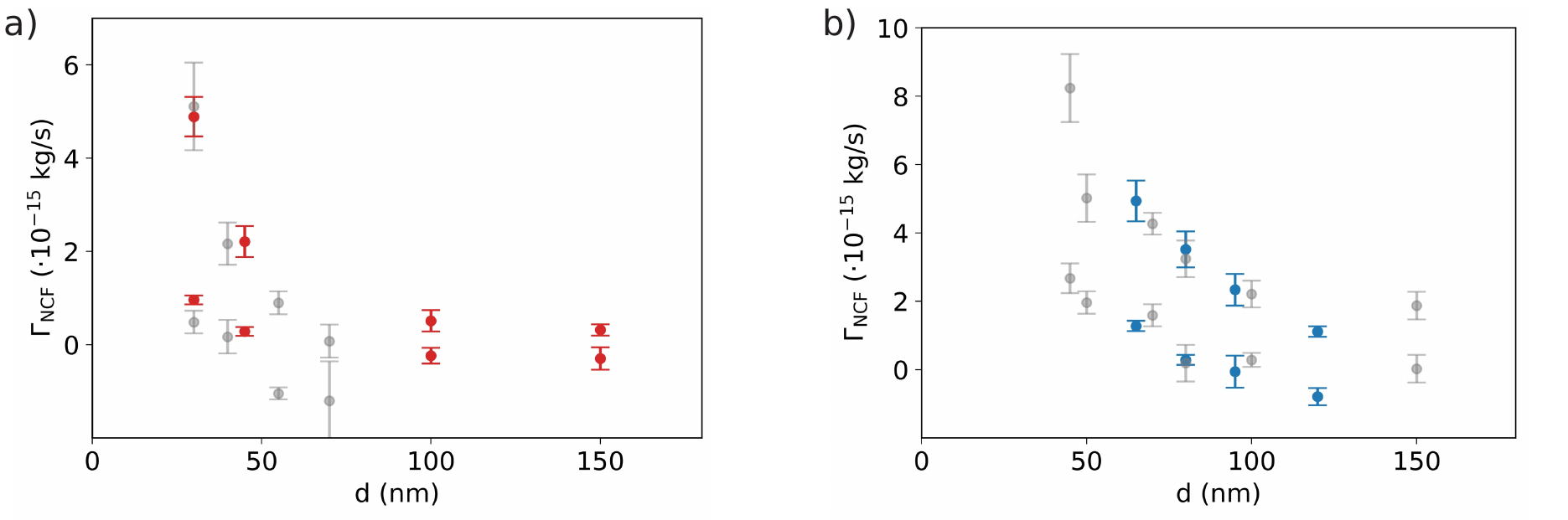}
	\caption{Maximum and minimum measured $\Gamma_\mathrm{NCF}$ under an external magnetic field of \SI{4}{\tesla} (grey data points). For comparison, the measured values without external field from Fig.~3 of the main manuscript are also shown (red, blue).}
	\label{fig:NCFmagneticfield}
\end{figure}

To test the surface dissipation added by the presence of the magnetic field, we extract the maxima and minima of the line scans of $\Gamma_\mathrm{NCF}$ in Fig.~\ref{fig:linescanmagnet}(c)-(d) and plot them versus $d$, see Fig.~\ref{fig:NCFmagneticfield}. For both materials, the data points agree with those for $B = 0$ within the expected statistical spread. From these findings, we conclude that the dominant contribution to $\Gamma_\mathrm{NCF}$ over both materials must be assigned to electrical fluctuations that are independent of $B$.

\begin{figure}[h!]
	\centering
\includegraphics{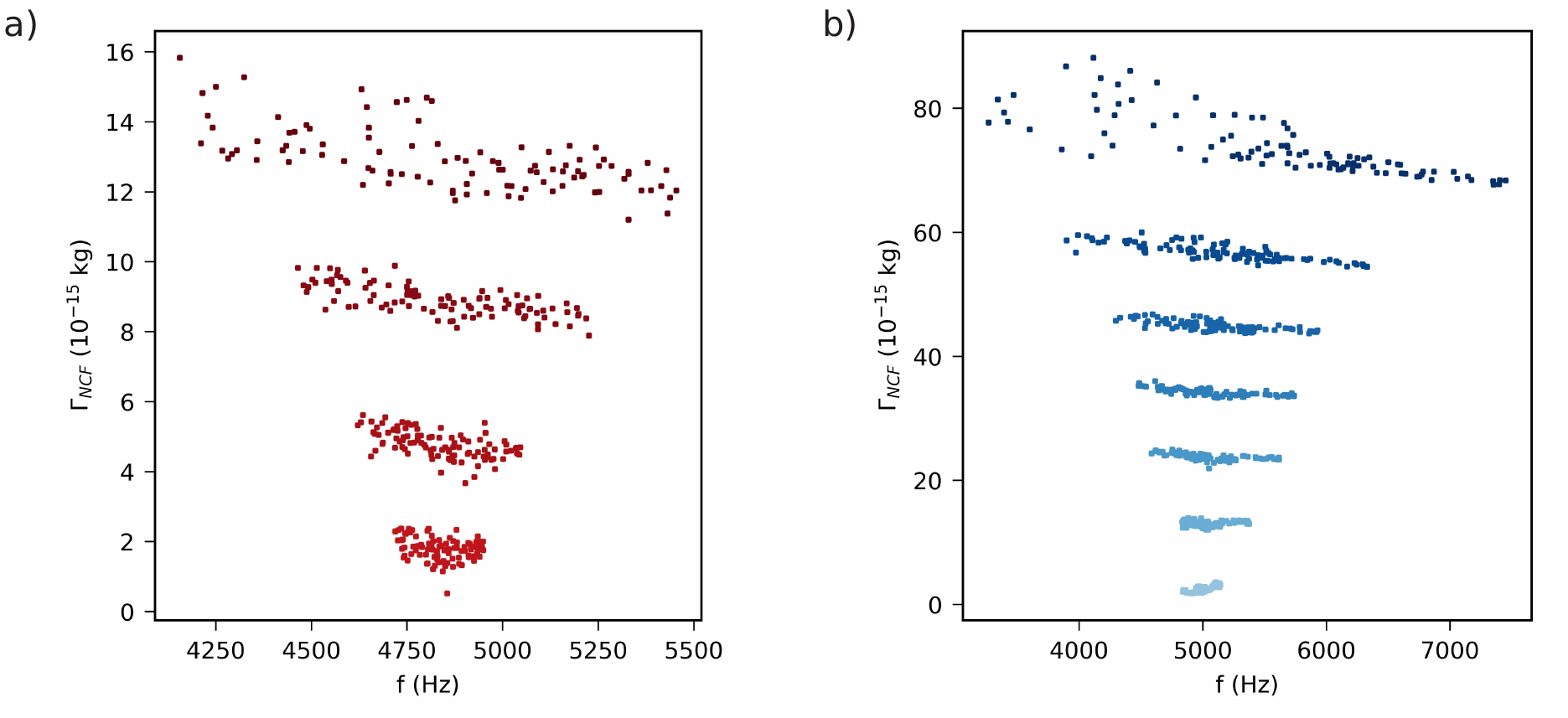}
	\caption{$\Gamma_\mathrm{NCF}$ as function of $f$ in the presence of an external magnetic field of $\SI{4}{\tesla}$ over Au (a) and over SiO\textsubscript{2} (b). Data are offset for better visibility by $\SI{3e-15}{\kilogram\per\second}$ each in (a), and $\SI{10e-15}{\kilogram\per\second}$ each in (b).}
	\label{fig:correlation}
\end{figure}

Finally, we show in Fig.~\ref{fig:correlation} the complementary plots to fig.~3(e)-(f) of the main text. As for $B = 0$, a clear correlation between $\Gamma_\mathrm{NCF}$ and $f$ is apparent. 

\clearpage

\section{Charge calibration}

\begin{figure}[h!]
	\centering
\includegraphics{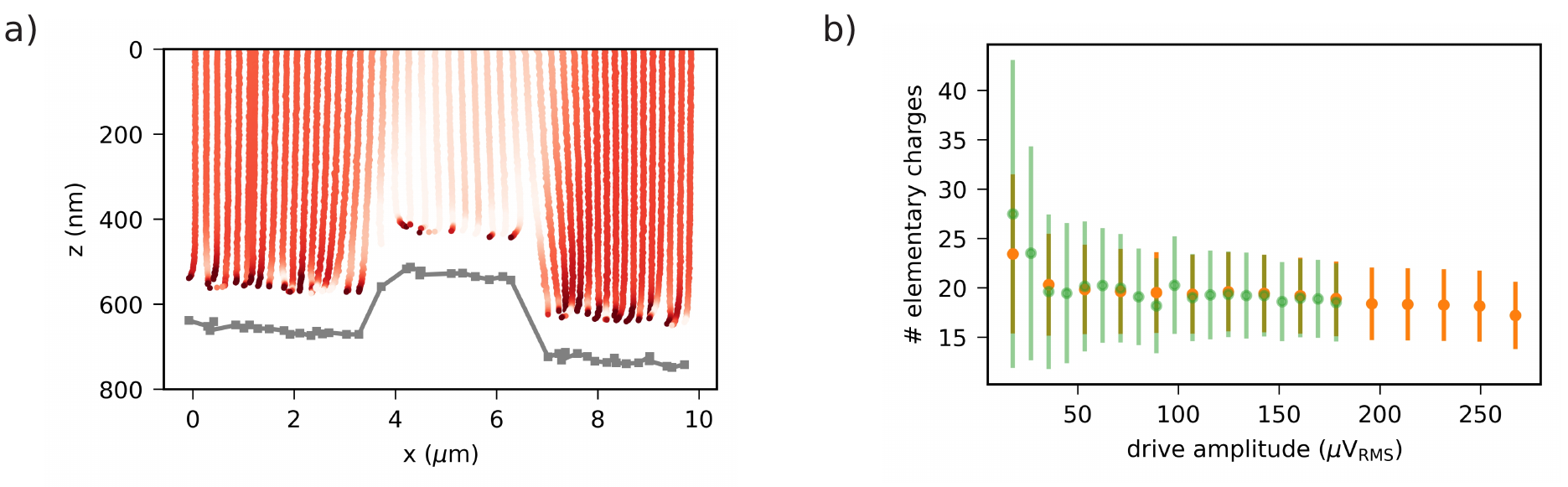}
	\caption{(a)~Cantilever static shift (lateral shift) and frequency (color coded: dark red $\SI{4.5}{\kilo \hertz}$, light red $\SI{5.5}{\kilo \hertz}$) for successive approaches above the sample. Grey points indicate the cantilever touch positions. The stripline profile is clearly visible. (b)~Results of the charge calibration. The orange and green data is measured at the positions $x = \SI{3.5}{\micro\meter}$ and $x = \SI{6.5}{\micro\meter}$ in (a), respectively.}
	\label{fig:chargecalibration}
\end{figure}

This section presents our calibration procedure to estimate the charge $q_\mathrm{tip}$ carried by the cantilever tip. We take advantage of the fact that our sample is a stripline used for inverting spins in magnetic resonance experiments~\cite{grob_magnetic_2019_s}. We apply a voltage to that stripline to electrostatically drive the cantilever vibrations. We then determine the force via the cantilever response function:
\begin{equation}
A(\omega) = \frac{F_d/m}{\sqrt{(\omega_0^2-\omega^2)^2+\left( \frac{\omega_0 \omega}{Q} \right)^2}}
\end{equation}where $A(\omega)$ is the cantilever amplitude, $F_d$ is the drive force, $m$ and $Q$ the cantilever's effective mass and quality factor, $\omega_0/2\pi = f_0$ its resonance frequency, and $\omega/2\pi$ the drive frequency. Measuring the amplitude at resonance, the force is therefore given by:
\begin{equation}
F_d = A(\omega_0) m \frac{\omega_0^2}{Q}.
\end{equation}
We simulate the electrical field $E_\mathrm{sim}(\textbf{r})$ at a position $\textbf{r}=(x,y,z)$ in COMSOL and use it to calculate the tip charge $q_\mathrm{tip}$ as
\begin{equation}
q_\mathrm{tip} = F_d(\textbf{r}) / E_\mathrm{sim}(\textbf{r})\,.
\end{equation}
Note that the cantilever position must be carefully determined. When hovering over the stripline, we measure lateral shifts of several $10-\SI{100}{\nano\meter}$ compared to its nominal position, see Fig.~\ref{fig:chargecalibration}(a). These shifts are monitored through changes in the feedback-controlled position of the cantilever in the interferometer fringe~\cite{grob_magnetic_2019_s}. We further found that the force values measured on both side of the stripline differ, which we ascribe to a potential difference between the cantilever tip and our setup ground. We managed to cancel this effect by applying a DC voltage of \SI{1.6}{\volt} to the stripline, resulting in a symmetric response and an estimated charge number of $q_\mathrm{tip}\approx 20$.


Our method overestimates the number of charges interacting with the sample surface during the scanning force microscopy experiments. The Coulomb electric field of a point charge decays as $1/d^2$ whereas the driving field generated by the $\approx \SI{2}{\micro\meter}$-broad stripline only starts to decrease as $1/d$ after a few micrometers. Our method therefore merely gives an upper boundary on the actual number of charges carried by the cantilever tip (within a few tens of nanometers from the apex).

\clearpage

\section{Electrostatic model}
\begin{figure}[h!]
	\centering
\includegraphics{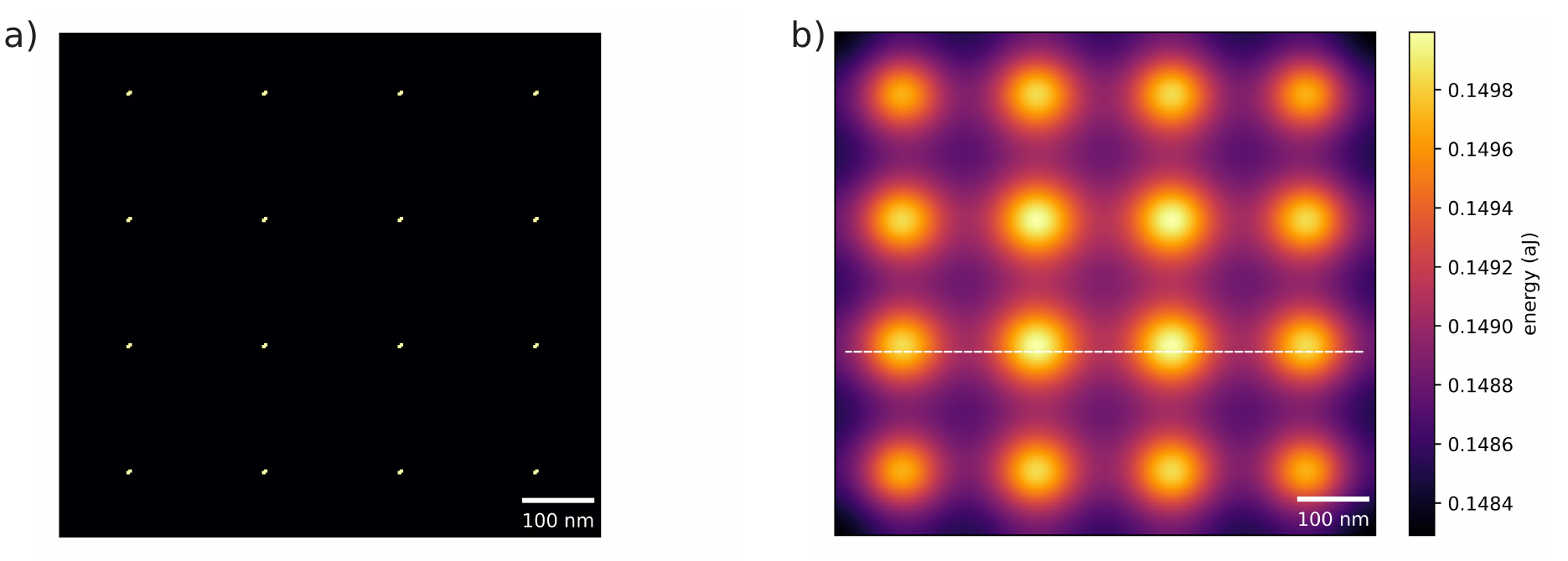}
	\caption{(a)~Simple charge density model. A charge $q_i$ is placed at each white dot. (b)~Energy map corresponding to $q_\mathrm{tip} = q_e$ and $q_i = 0.7 q_e$ at a surface-charge distance $d+\Delta = \SI{50}{\nano\meter}$. The dashed line shows the position of the line scan in Fig.~\ref{fig:linescansimuAlex}.}
	\label{fig:energymap}
\end{figure}

\begin{figure}[h!]
	\includegraphics{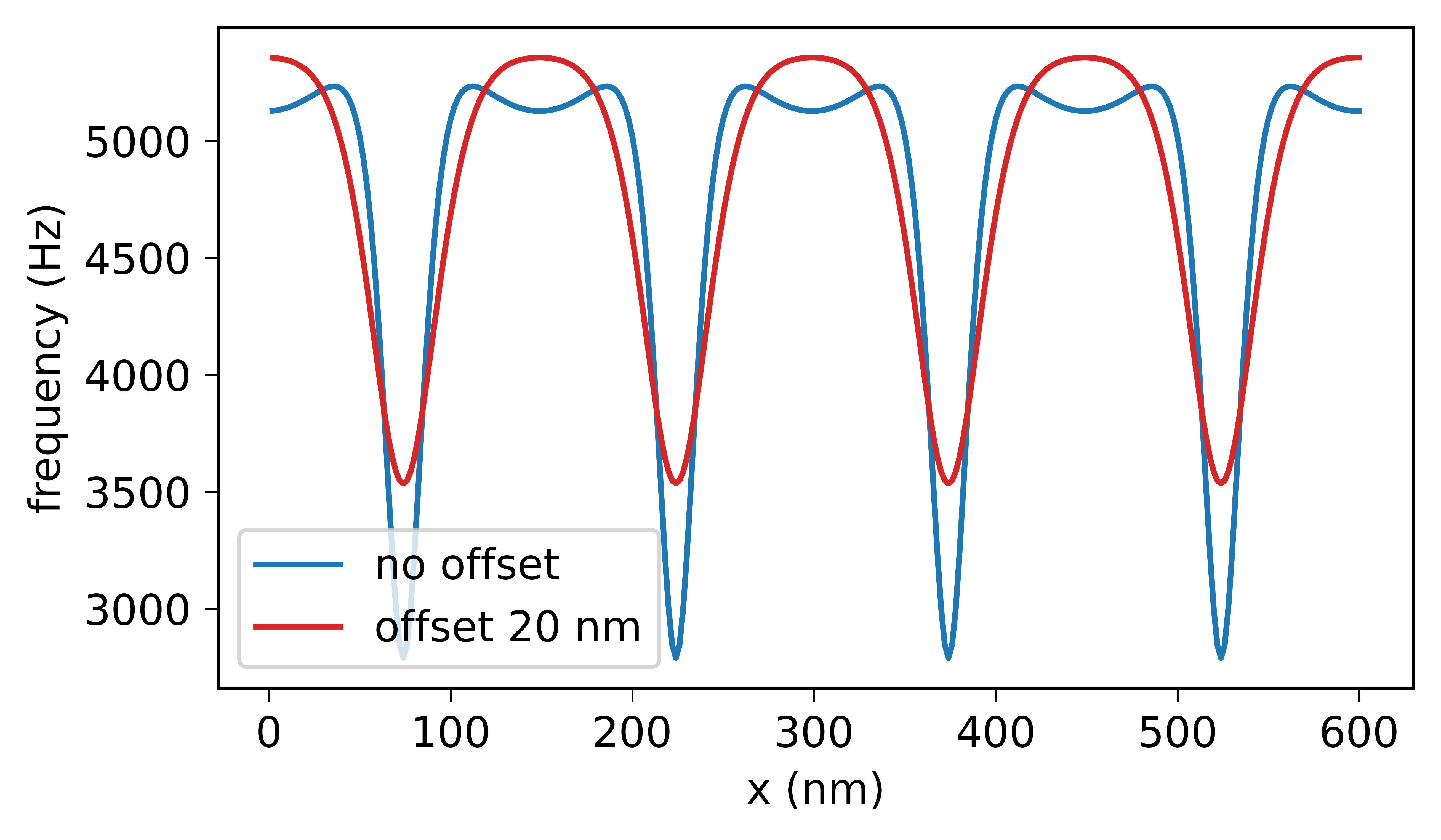}
	\caption{Calculated frequency along the white dashed line in Fig.~\ref{fig:energymap}(b). The result for $d = \SI{30}{\nano\meter}$ without an offset is shown in blue, yielding a large asymmetry between the positive and negative frequency deviations and characteristic double maxima between charge sites that we do not observe in the experiment, cf. Fig~2(c) of the main text. Including a tip-charge offset of $\Delta = \SI{20}{\nano\meter}$ results in a much more realistic simulation.}
	\label{fig:linescansimuAlex}
\end{figure}

In the main text, we explain the simple electrostatic model used to calculate the variable frequency
\begin{align}
    f = \frac{1}{2\pi} \sqrt{\frac{k_\mathrm{0}}{m} + \frac{k_{\mathrm{el}}}{m}}\,,
\end{align}
where $k_{\mathrm{el}} = \delta^2 E_{\mathrm{el}}/\delta x^2$ is the electrical spring constant obtained from the summed Coulomb energy of charges $q_i$ on the surface. For the model comparison in Fig.~3, we placed charges on a grid with a separation of \SI{150}{\nano\meter}, see Fig.~\ref{fig:energymap}. To avoid edge effects, we simulate a $10\times 10$ times larger area than that shown (i.e., evaluated). Two free parameters are involved in this simulation: first, we fix the value of $q_\mathrm{tip} q_i = 0.7 q_e^2$ by comparing the maximum and minimum frequencies found in a simulated line scan, see Fig.~\ref{fig:linescansimuAlex}. Second, the characteristic asymmetry observed for $d = \SI{30}{\nano\meter}$ over Au provides us with a criterion to estimate the offset $\Delta$ between the tip apex and the effective charge position. For $\Delta = 0$, the model corresponds to a charge placed at the very apex, and the simulated line scan produces characteristic double peaks that are clearly absent in the experiment, see blue trace in Fig.~\ref{fig:linescansimuAlex}. An offset of $\Delta = \SI{20}{\nano\meter}$ ($\pm\SI{5}{\nano\meter}$) removes these double peaks, while preserving the asymmetry between the sharp dips and the broad maxima, see red trace in Fig.~\ref{fig:linescansimuAlex}. The asymmetry stems from the difference between placing the tip directly over a repulsive charge (frequency dip) or at a position between two charges (frequency maximum). The offset is comparable to the nanowire tip radius of $\sim\SI{10}{\nano\meter}$ reported in an earlier publication~\cite{tao_single_2015_s}.

\section{Surface topography images}

\begin{figure}[h!]
	\centering
\includegraphics{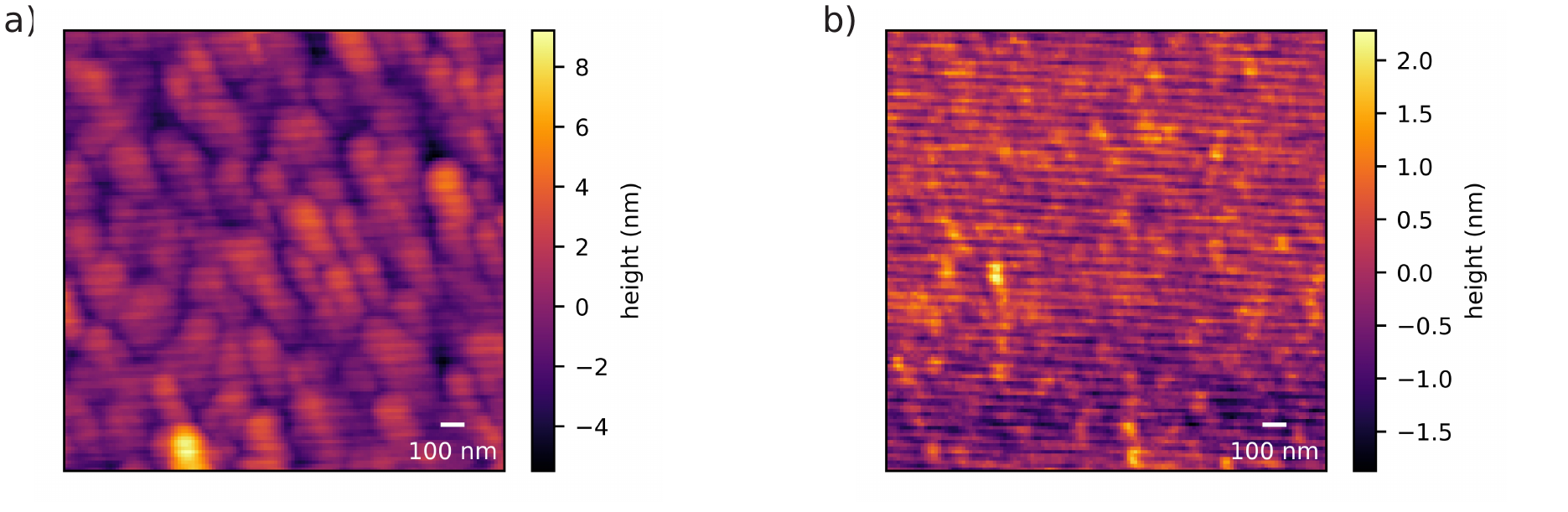}
	\caption{Surface topography of (a)~Au and (b)~SiO\textsubscript{2} surfaces on a second sample chip, measured with a commercial AFM.}
	\label{fig:afmmaps}
\end{figure}

In Fig.~\ref{fig:afmmaps}, we display the surface topography of Au and SiO\textsubscript{2} on a second sample chip. The chip is cleaved from the same thermally oxidized Si wafer, and the Au layer was evaporated in the same process as for the sample mounted in out nanoladder scanning force microscope. The topography images in Fig.~\ref{fig:afmmaps} were measured with a commercial atomic force microscope (AFM). They clearly show difference in topography between the two surfaces. In particular, the lateral grain sizes on Au are about $100-\SI{150}{\nano\meter}$ with a typical height of $4-\SI{6}{\nano\meter}$, while those on SiO\textsubscript{2} appear to be smaller, roughly $20-\SI{50}{\nano\meter}$ laterally and $1-\SI{1.5}{\nano\meter}$ high. From these differences, we conclude it unlikely that the very similar frequency maps in Fig.~2(a)-(b) of the main text are directly caused by topographic variations.

\section{Frequency map over a different sample}
\begin{figure}[h!]
	\includegraphics{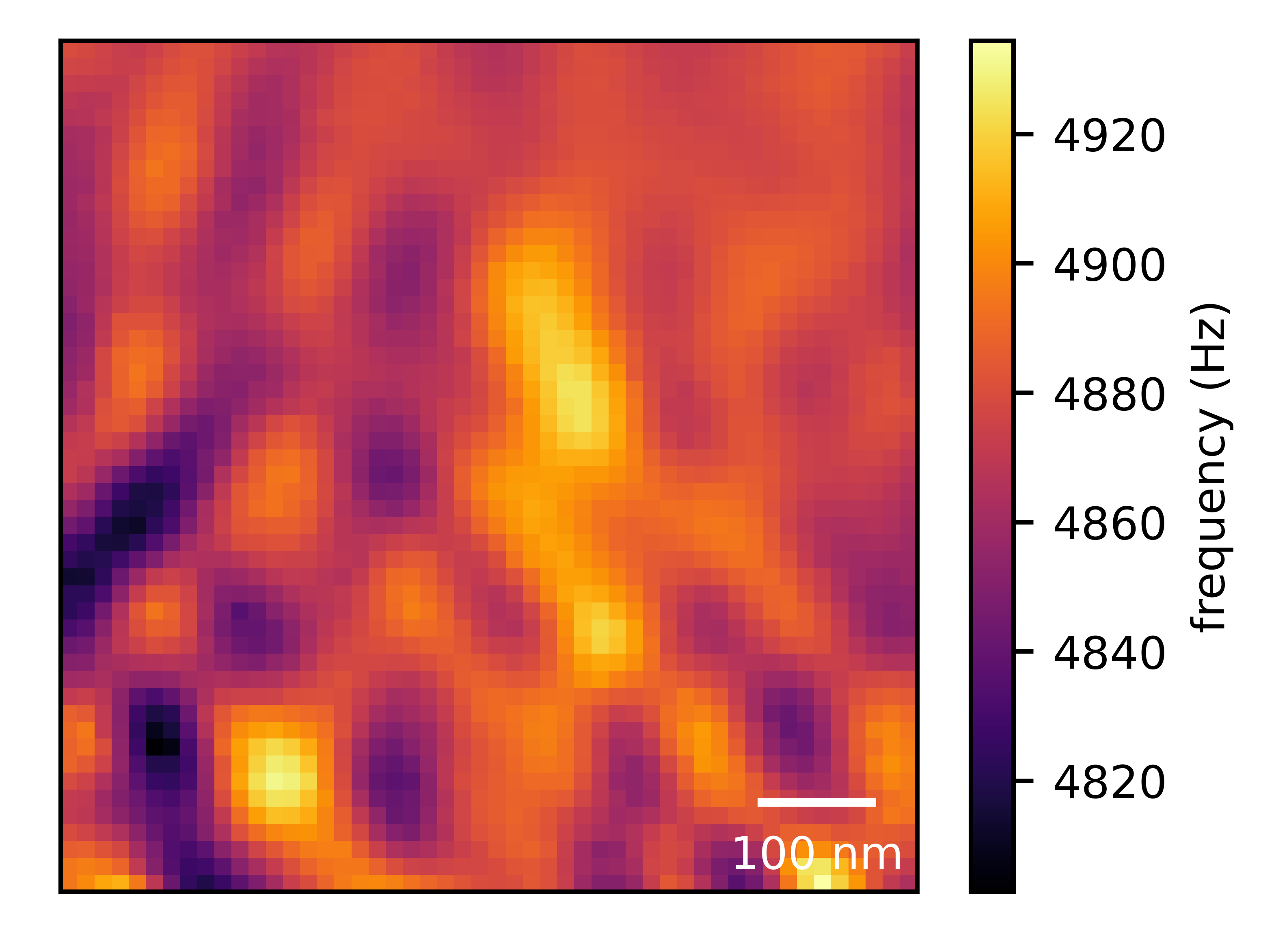}
	\caption{Frequency map measured $\SI{50}{\nano \meter}$ over Pt.}
	\label{fig:WHmap}
\end{figure}

In a separate experiment, we measured the cantilever frequency $f$ over a $\SI{3}{\nano\meter}$-thick Pt layer that was E-beam evaporated over diamond-like carbon, see Fig.~\ref{fig:WHmap}. We observe a similar pattern as on Au and SiO\textsubscript{2}, suggesting the presence of potential patches. We did not study the non-contact friction over this sample.

\section{XPS sample analysis}
\begin{figure}[h]
	\centering
\includegraphics{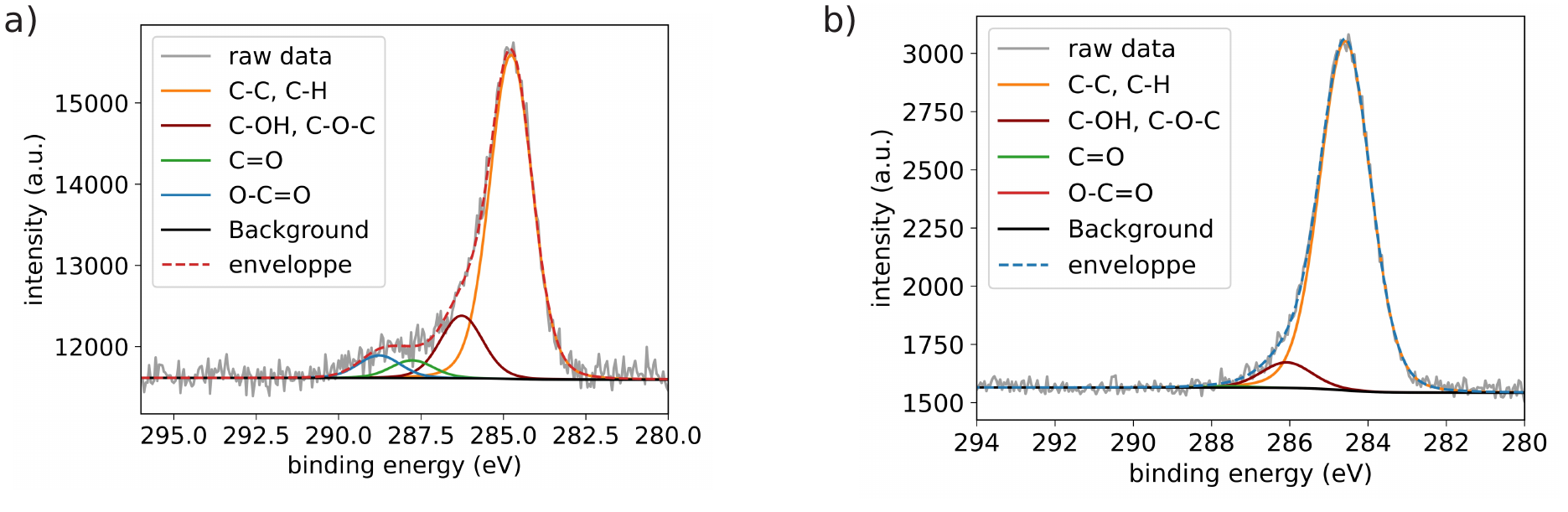}
	\caption{Representative X-ray photoelectron spectra of (a)~Au and (b)~SiO\textsubscript{2} surfaces showing high-resolution C 1s regions. The shape of the spectra indicate that the composition of adsorbed hydrocarbon contamination differ between Au and SiO\textsubscript{2} surfaces. Specifically, deconvolution of the C 1s signals suggests a greater contribution of C-O bonding character in the thin layers of adventitious carbon material present on Au compared to SiO\textsubscript{2}. The non-identical nature of the adsorbed material can lead to different dielectric parameters due to the additional interaction of permanent dipole moments in polar, oxygen-containing species with electric fields~\cite{Sen_1992}.}
	\label{fig:xps}
\end{figure}

We performed X-ray photoelectron spectroscopy using a PHI Quantera SXM photoelectron spectrometer at the Swiss Federal Laboratories for Materials Science and Technology (EMPA), see Fig.~\ref{fig:xps}. A monochromatic Al K$_\alpha$X-ray source with a $\SI{100}{\micro\meter}$ circular spot size was used under ultrahigh vacuum ($\SI{1e-9}{\milli\bar}$). High-resolution C 1s spectra were acquired at a pass energy of $\SI{55}{\eV}$ using a $\SI{20}{\milli\second}$ dwell time. For all scans, $\SI{15}{\kilo\volt}$ was applied with an emission current of $\SI{3}{\milli\ampere}$; an average of 8-10 scans were collected per region. Spectra were fit with CasaXPS Software Version 2.3.23PR1.0 using a Shirley background and Gaussian-Lorenzian peak shapes to deconvolute contributions to the adventitious carbon signals. The dominant peaks were assigned to sp3 carbon and calibrated to $\SI{284.8}{\eV}$ as a charge reference; satellite peaks at higher binding energies of ca. $+\SI{1.5}{\eV}$ and circa $+3-\SI{4}{\eV}$ were assigned to carbon singly and doubly bound to oxygen, respectively~\cite{Sangtawesin2019}. Relative signal contributions were averaged over spectra collected from three distinct regions for both Au and SiO\textsubscript{2} surfaces.

\clearpage

\let\oldthebibliography=\thebibliography
\let\oldendthebibliography=\endthebibliography
\renewenvironment{thebibliography}[1]
     {\section*{\refname}%
      \@mkboth{\MakeUppercase\refname}{\MakeUppercase\refname}%
      \list{}%
           {\setlength{\labelwidth}{0pt}%
            \setlength{\labelsep}{0pt}%
            \setlength{\leftmargin}{\parindent}%
            \setlength{\itemindent}{-\parindent}%
            \@openbib@code
            \usecounter{enumiv}				%
					  \oldthebibliography{#1}%
            \setcounter{enumiv}{0}
								{\oldendthebibliography}}%
      \sloppy
      \clubpenalty4000
      \@clubpenalty \clubpenalty
      \widowpenalty4000%
      \sfcode`\.\@m}
     {\def\@noitemerr
       {\@latex@warning{Empty `thebibliography' environment}}%
      \endlist}
\makeatother
\renewcommand{\bibsection}{}

\end{document}